\documentclass[fleqn,usenatbib]{mnras}

\usepackage[english]{babel}
\usepackage{amsmath,amssymb}
\usepackage{graphicx, subfig}
\usepackage[normalem]{ulem}

\usepackage{verbatim}

\usepackage{color}
\usepackage{multirow}
\usepackage{mathtools}
\usepackage{epstopdf}

\usepackage{url}
\usepackage{xcolor}

\graphicspath{{figures/}{../figures/}}

\usepackage{etoolbox}
\makeatletter
\patchcmd\@combinedblfloats{\box\@outputbox}{\unvbox\@outputbox}{}{\errmessage{\noexpand patch failed}}
\makeatother

\def\nat{Nature}

\def\mnras{MNRAS}
\def\apj{ApJ}
\def\apjl{ApJL}
\def\apjs{ApJS}
\def\aap{A\&A}
\def\aaps{A\&A Supp.}

\def\araa{ARA\&A}
\def\aj{AJ}

\def\pasa{Publ. Astr. Soc. Australia}
\def\pasp{Publ. Astr. Soc. Pac.}

\def\be{\begin{equation}} 
\def\ee{\end{equation}} 
\def\ba{\begin{eqnarray}} 
\def\ea{\end{eqnarray}}

\def\kms{\,{\rm {km\, s^{-1}}}} 
 
\def\msun{{\Msun}}

\def\HH{${\rm {H_2}}$}

\def\gsim{\lower.5ex\hbox{\gtsima}} 
\def\lsim{\lower.5ex\hbox{\ltsima}} \def\gtsima{$\; \buildrel > \over 
\sim \;$} \def\ltsima{$\; \buildrel < \over \sim \;$} \def\prosima{$\; 
\buildrel \propto \over \sim \;$} \def\gsim{\lower.5ex\hbox{\gtsima}} 
\def\lsim{\lower.5ex\hbox{\ltsima}} 
\def\simgt{\lower.5ex\hbox{\gtsima}} 
\def\simlt{\lower.5ex\hbox{\ltsima}} 
\def\simpr{\lower.5ex\hbox{\prosima}} \def\la{\lsim} \def\ga{\gsim} 
  
 \def\gtsima{$\; \buildrel > \over \sim \;$} 
\def\ltsima{$\; \buildrel < \over \sim \;$} 
\def\gsim{\lower.5ex\hbox{\gtsima}} 
\def\lsim{\lower.5ex\hbox{\ltsima}} 
\def\simgt{\lower.5ex\hbox{\gtsima}} 
\def\simlt{\lower.5ex\hbox{\ltsima}} 
\def\simpr{\lower.5ex\hbox{\prosima}} 
\def\la{\lsim} 
\def\ga{\gsim}

\def\msun{\,{\rm \Msun}}

\def\E3{{\cal E}_{\rm g}^{III}}

\def\Zsun{\rm Z_\odot}

\def\r12{r_{1/2}} 
\def\x12{x_{1/2}} 
\def\v12{v_{1/2}}

\newcommand\code[1]{\textsc{\MakeLowercase{#1}}}

\def\nh2{n_{\rm H2}}
\def\fh2{f_{\rm H2}}

\def\arcsec{^{\prime\prime}}
\def\angstrom{\textrm{A\kern -1.3ex\raisebox{0.6ex}{$^\circ$}}}
\def\myr{\rm Myr}

\def\msun{{\rm M}_{\odot}}
\def\zsun{{\rm Z}_{\odot}}

\def\msunyr{\msun\,{\rm yr}^{-1}}

\def\voldpc{\msun\,{\rm pc}^{-3}}
\def\kpc{{\rm kpc}}

\def\althaea{Alth{\ae}a}
\def\highz{$\mbox{high-}z$~}

\makeatletter
\def\@hex@@Hex#1%
 {\if a#1A\else \if b#1B\else \if c#1C\else \if d#1D\else
  \if e#1E\else \if f#1F\else #1\fi\fi\fi\fi\fi\fi \@hex@Hex}
\makeatother

\definecolor{apcolor}{HTML}{b3003b}
\definecolor{cbcolor}{HTML}{ff0f00}
\definecolor{afcolor}{HTML}{b3443c}
\definecolor{ddcolor}{HTML}{077a2f}
\definecolor{vgcolor}{HTML}{8F00FF}

\definecolor{sscolor}{HTML}{e0bf06}

\title[Stars of \highz dwarf galaxies]{The stellar populations of high-redshift dwarf galaxies}

\author[V. Gelli et al.]{
V. Gelli$^{1,2}$\thanks{E-mail: \href{mailto:viola.gelli@unifi.it}{viola.gelli@unifi.it}},
S. Salvadori$^{1,2}$,
A. Pallottini$^{3,4}$,
A. Ferrara$^{4,5}$
\\
$^{1}$ Dipartimento di Fisica e Astronomia, Universit\'{a} degli Studi di Firenze, via G. Sansone 1, Sesto Fiorentino, Italy\\
$^{2}$ INAF/Osservatorio Astrofisico di Arcetri, Largo E. Fermi 5, Firenze, Italy\\
$^{3}$ Centro Fermi, Museo Storico della Fisica e Centro Studi e Ricerche ``Enrico Fermi'', Piazza del Viminale 1, Roma, 00184, Italy \\
$^{4}$ Scuola Normale Superiore, Piazza dei Cavalieri 7, I-56126 Pisa, Italy\\
$^{5}$ Kavli Institute for the Physics and Mathematics of the Universe (WPI), University of Tokyo, Kashiwa 277-8583, Japan\\
}

\date{Accepted XXX. Received YYY; in original form ZZZ}

\pubyear{2020}

\begin{document}
\label{firstpage}
\pagerange{\pageref{firstpage}--\pageref{lastpage}}
\maketitle

\begin{abstract}
We use high-resolution ($\approx 10$ pc), zoom-in simulations of a typical (stellar mass $M_\star\simeq10^{10}\msun$)  Lyman Break Galaxy (LBG) at $z\simeq 6$ to investigate the stellar populations of its six dwarf galaxy satellites, whose
stellar [gas] masses are in the range $\log (M_\star/\msun) \simeq 6-9$  [$\log (M_{gas}/\msun) \simeq4.3-7.75$].
The properties and evolution of satellites show no dependence on the distance from the central massive LBG ($< 11.5$~kpc). Instead, their star formation and chemical enrichment histories are tightly connected their stellar (and sub-halo) mass.
High-mass dwarf galaxies ($\rm M_\star \gtrsim 5\times 10^8 \msun$) experience a long history of star formation, characterised by many merger events. 
Lower-mass systems go through a series of short star formation episodes, with no signs of mergers; their star formation activity starts relatively late ($z\approx 7$), and it is rapidly quenched by internal stellar feedback.
In spite of the different evolutionary patterns, all satellites show a spherical morphology, with ancient and more metal-poor stars located towards the inner regions. All six dwarf satellites experienced high star formation rate ($\rm >5\,\msunyr$) bursts, which can be detected by JWST while targeting high-$z$ LBGs.
\end{abstract}

\begin{keywords}
galaxies: dwarf, high-redshift, formation, evolution -- cosmology: theory
\end{keywords}




\section{Introduction}

Dwarf galaxies (stellar mass $\rm M_\star \lesssim 10^9 \msun$) are the most common type of galaxies within the Local Group, where over 70 of such systems are observed as satellites of the Milky Way (MW) and Andromeda galaxies \citep{Tolstoy09, McCon12}.
According to the hierarchical $\rm \Lambda CDM$ model, these small systems are expected to be the most abundant at all cosmic times. They are the first galaxies to form in the Universe; they also play a fundamental role during the early epochs of structure formation for several reasons. For example, they represent the building blocks of more massive galaxies like the MW. Second, they hosted the first generation of stars marking the end of the Dark Ages and hence the beginning of the Epoch of Reionization (EoR, $z>6$). Third, because of their shallow potential wells, they started to pollute the pristine intergalactic medium (IGM) with newly produced heavy elements, injected both via stellar winds and supernova (SN) explosions. 
Finally,  \highz dwarf galaxies are also believed to be the main contributors to the process of reionization as suggested by numerous studies based on semi-analitical models \citep[e.g.][]{Salvadori14,Yue16}, numerical simulations \citep[e.g.][]{Wise14,Rosdahl2018}, and observations \citep[e.g.][]{Bowens12, Finkelstein16}. 

On the other hand, the same feedback processes that make dwarf galaxies the key drivers of the early cosmic evolution, strongly affect their formation and evolution. In particular, both UV radiation from massive stars, and the mechanical energy injected by SNe heat the gas,  leading to a temporary or permanent suppression of the star formation (SF) in these small systems. The fact that dwarf galaxies are very sensitive to these negative feedback processes, as well as possible gravitational interactions, makes them the ideal laboratories to study these effects during early galaxy formation. 

Cosmological hydrodynamical simulations have lately proven to be an invaluable tool for the study of galaxy evolution. They have convincingly shown that the inclusion of detailed feedback effects is essential to reproduce the evolution of the IGM and the \highz  \citep[e.g.][]{Gnedin14,Pallottini14} and present-day \citep[e.g.][]{illustris,Schaller2015} structures. 
We are now at a stage where cosmological simulations are detailed enough as to allow a thorough study of the inner structure and properties of even the smallest systems, i.e. dwarf galaxies, and a meaningful comparison with observational data. Thanks to the zoom-in technique it is possible to reach extremely high spatial and mass resolutions and a great effort has already been made in this direction to explore the formation of dwarf galaxies in various environments and at different cosmic times. This effort includes simulations of local ($z=0$) dwarf galaxies orbiting around the Milky Way \citep[e.g.][]{RevazJablonka18,Garrison-Kimmel19}, of field dwarf galaxies in the local Universe \citep[e.g.][]{lupi:2020}, during the epoch of reionization \citep[e.g.][]{Ma18}, and of the earliest star forming dwarf galaxies hosting the first stars \citep[e.g.][]{JeonBromm19}. Noticeably -- to our knowledge -- a study of high-z dwarf galaxies located in dense cosmic environments has not been performed yet.

From an observational point of view, many nearby faint dwarf galaxies have been detected and analysed in detail \citep[e.g.][]{Tolstoy09}. Still, due to the enormous variety of these systems, even in the extensively studied Local Group many issues and questions regarding their formation and evolution remain unanswered: which processes drive the variegated star formation histories observed? How many of these small systems contributed to the build-up of the central Milky Way galaxy?

At high redshift, dwarf galaxies pose an even greater challenge since they appear to be of fundamental importance in early cosmological scenarios, but no observations are currently available. However, our knowledge of the high-$z$ Universe will be largely improved with the advent of the {\it James Webb Space Telescope} (JWST, to be launched in March 2021). JWST will be able to investigate the faint-end slope of the galaxy UV luminosity function \citep[e.g.][]{Finkelstein16}. Amongst its targets, JWST will observe many high redshift Lyman Break Galaxies (LBG). Such typically massive and star forming systems are usually located in high-density cosmic environments; theory also predicts that many dwarf galaxies should be present in their vicinity as well, most likely as satellites. Thus, it is timely to turn our attention to the characterisation of these important low-mass systems.

In this paper we concentrate on (i) the expected stellar properties of these objects, and (ii) how they are influenced by feedback effects, (iii) and by the presence of a nearby massive host galaxy. In order to achieve this objective, we use state-of-art, high resolution zoom-in simulations of a typical $z\simeq 6$ LBG \citet{Pallottini17}.

In Sec. \ref{sec_simulation} we briefly present the main features of the hydro-simulations, focusing particularly on the identification of the satellite galaxies. In Sec. \ref{sec_properties} we illustrate the main properties of dwarf satellites at $z = 6$ (density, radial profiles and stellar metallicity); Sec. \ref{formationandevolution} analyses their star formation and metal enrichment histories. In Sec. \ref{sec_interpretation} we interpret the above results and investigate how different environmental and feedback effects regulate the predicted  evolution. We discuss global galaxy properties in Sec. \ref{sec_discussion}, and finally draw our conclusions in Sec. \ref{sec_conclusions}.

\section{Cosmological zoom-in simulations}\label{sec_simulation}

\begin{figure*}
\centering
\includegraphics[width=0.95\textwidth]{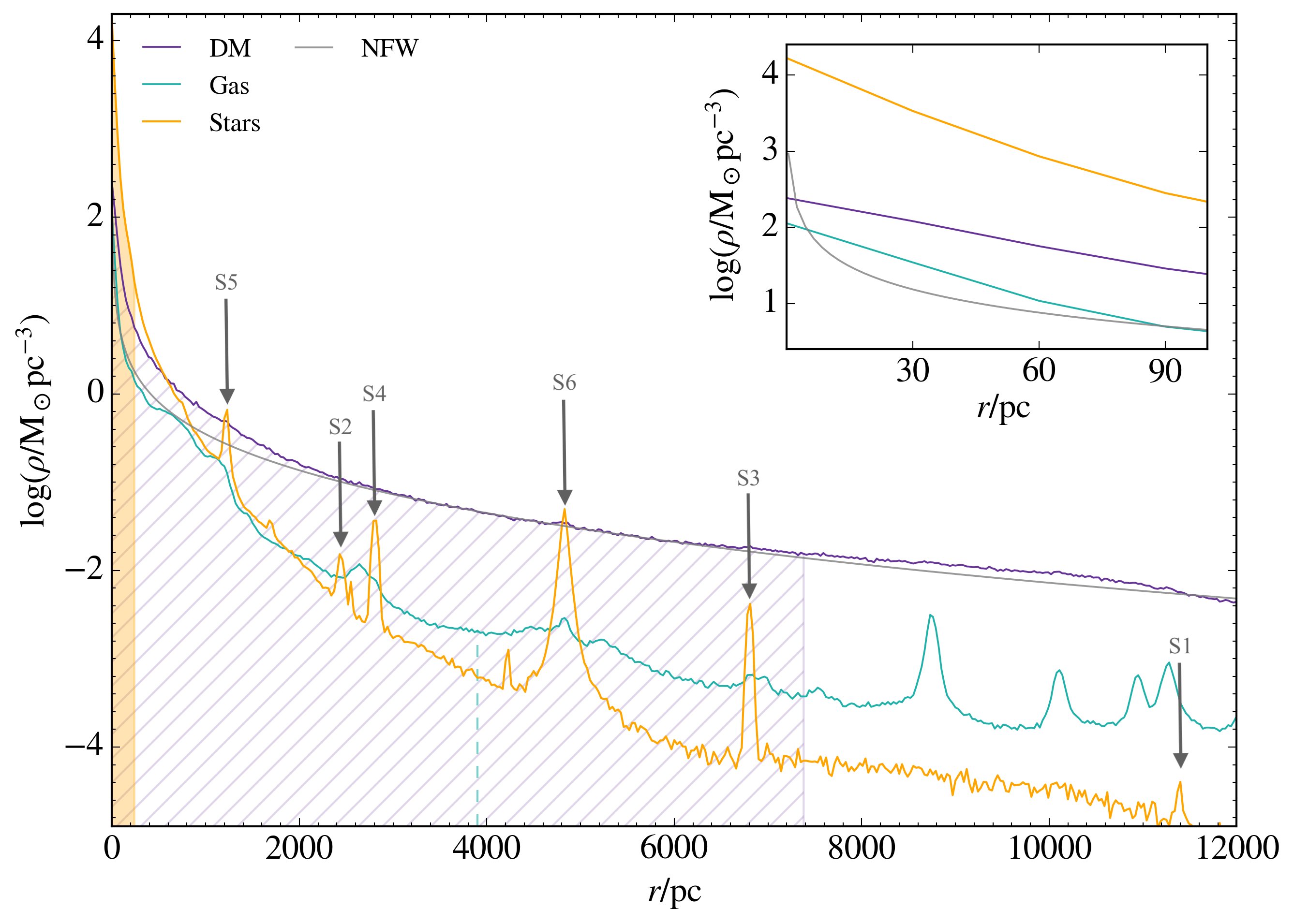}
\caption{Spherically-averaged mass density profiles for DM, stars, and gas in \althaea. The shaded orange area, the dashed green line and the grey hatched area mark regions enclosing 50\% of the stellar, gas and DM mass, respectively, contained in a volume of size 20~kpc centred on \althaea. The inset is a zoomed view of the inner 100 pc.
\label{fig:rad_althaea}
}
\end{figure*}

In order to analyse the properties of the dwarf satellites of a typical \highz Lyman Break Galaxy, we use the high resolution cosmological simulation described in \cite{Pallottini17}. It uses a customised version of the adaptive mesh refinement code RAMSES \citep{Teyssier02} to follow the evolution of a dark matter halo and its environment with a zoom-in technique. This allows to resolve spatial scales of $\simlt$ 30 pc, with a baryon (DM) mass of $\simeq 2\times 10^4\msun$ ($\simeq 10^5\msun$). The target halo has a mass $M_{h}\simeq 3\times 10^{11}\msun$ at $z\simeq6$ (virial radius $r_{vir}\simeq 26\, \kpc$) and hosts ``\althaea", a proto-typical LBG given its stellar mass $M_\star\simeq10^{10}\msun$, star formation rate ${\rm SFR}\simeq 100\,\msunyr$, and a morphology characterised by a spiral disk of radius $\simeq 0.5$ kpc.

The multi-scale initial condition are set with \code{music} \cite[][]{HahnAbel11} at $z=100$, and the gas is set to a non-zero metallicity floor $Z_{floor}=10^{-3}\zsun$ since our resolution is not sufficient to follow the formation of the first stars in pristine mini-halos \citep[e.g.][]{Wise12,OShea15}. This value is in agreement with the metallicity of the diffuse IGM found in cosmological simulations of metal enrichment \citep[e.g.][]{Dave11,pallottini:2014b,MaioTescari15}, and it very marginally affects the gas cooling time. In the simulation stars form out of gas according to the Schmidt-Kennicutt relation \citep{Kennicutt98} depending on the molecular hydrogen fraction, which is computed by adopting a non-equilibrium chemical network implemented via the \code{KROME} code \citep{krome}. Stellar feedback includes supernovae, winds from massive stars and radiation pressure \citep{Pallottini16}. In particular the energy inputs and chemical yields depend on the age of the stellar population and are computed for different metallicities with \code{STARBURST99}\footnote{The stellar metallicity values covered by STARBURST99 are: $Z_\star / \Zsun = 0.02, 0.2, 0.4, 1$.} \citep{starburst99} using PADOVA stellar tracks \citep{Bertelli94} and a Kroupa initial mass function \citep{Kroupa01}. 
This approach is similar to the one typically adopted in simulations which have a better spatial and time resolution \citep[e.g.][]{Kim14,lupi:2020}. Mechanical feedback from SNe is taken into account through a detailed modelling of the unresolved physics inside molecular clouds, considering the blastwave propagation through its different evolutionary stages \citep[based on][]{OstrikerMcKee98}. The model is in broad agreement with more specific numerical studied \citep[e.g.][]{Martizzi15,WalchNaab15}.
The turbulent and thermal energy content of the gas is modelled as in \cite{Agertz13}.  The interstellar radiation field (ISRF) is approximated as spatially uniform and its intensity scales with the SFR of \althaea~\citep[for a discussion, see][]{Pallottini19}.

In Fig. \ref{fig:rad_althaea} we plot the spherically-averaged radial density profiles for DM, stars and gas in the inner $\simeq 12\,\rm kpc$ from the centre of Althea (inner 100 pc in the inset). The stellar contribution dominates in the inner regions where the density reaches $\simeq 10^4 \voldpc$. The gas density smoothly decreases from the central peak of $\simeq 100 \voldpc$, but it exceeds the stellar density for $r\gtrsim2\,\rm \kpc$. At distances $\simgt 500 \, \rm pc$ the DM becomes the dominant component.
The grey curve represents the Navarro-Frenk-White (NFW) density profile \citep{NFW}, obtained assuming a concentration parameter of $c=4$ \citep[e.g.][]{Angel16}. Comparing it with the DM, we see that the two profiles are consistent at the scale radius $r_s = r_{vir}/c \simeq 6.5 \, \rm kpc$ while in the inner regions the NFW features the classical cusp, which however is not present in the simulated halo. The absence of a cusp is actually in agreement with observations of DM profiles in local dwarf galaxies \citep[e.g.][]{moore:1994Natur,walker:2011ApJ}. While the absence of a cusp might have a physical origin (e.g. SN feedback \citep[e.g.][]{Pontzen12}), the spatial resolution of our simulation, although very high,  is probably not yet sufficient to draw a firm conclusion on this issue, which in any case is beyond the scope of our study.

The prominent peaks of the stellar and gas radial profiles in Fig. \ref{fig:rad_althaea} correspond to our target satellite structures. Not surprisingly, these enhancements are not seen in the DM component as \althaea's DM halo preserves a high density value ($\rho_{\rm DM}\gtrsim10^{-2} \voldpc$) up to $r_{vir}$. Although, as we will see in Sec. \ref{radialprofiles}, the satellites have a central DM density of $10<\rho_{\rm DM}/ \voldpc<10^{2}$, their signal is washed away by the spherical average. For similar reasons, it is difficult to identify dwarf satellites by using clustering algorithms based on local overdensities or distance-based Friends-of-Friends algorithms \citep[e.g.][]{springel2001,knollmann2009,behroozi:2013,elahi:2019}. 

Because of the limited number of objects, here it is more effective (and simpler) to locate satellites directly through their stellar (or gas) component, as described in the following paragraph. We then check \textit{a posteriori} the presence of DM to confirm the identification. For example, when studying the DM profiles of satellite dwarf galaxies (Sec. \ref{radialprofiles}), we would expect their DM to have a central peak and then flatten at a floor value corresponding to the halo DM density (Fig. \ref{fig:rad_althaea}) at the location of the satellite. 

\begin{figure*}
	\centering
	\includegraphics[width=0.93\textwidth]{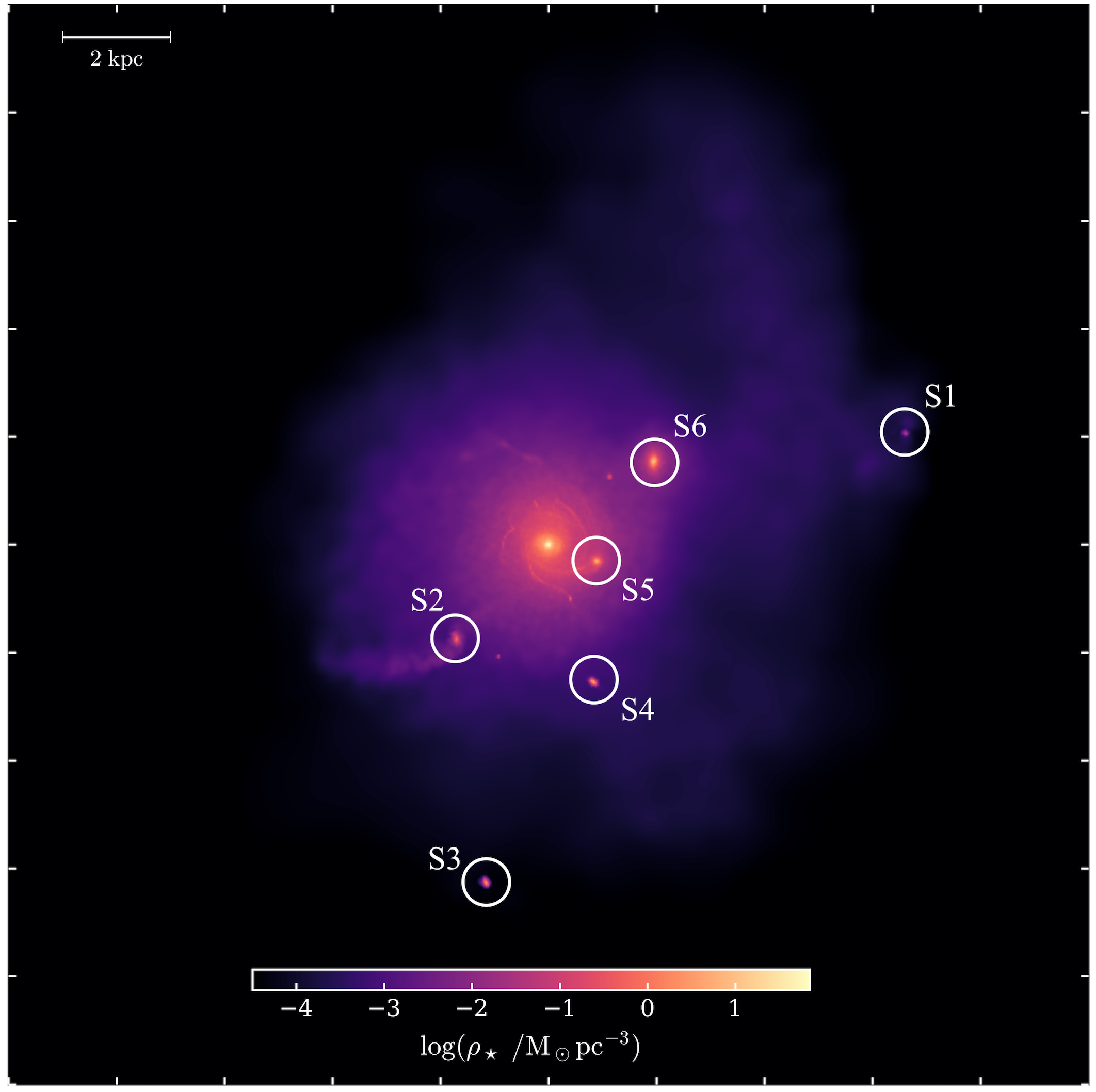}
    \caption{
    Stellar map centred on the simulated galaxy \althaea\, at $z=6$. The field of view is 20 kpc$\times$ 20 kpc (corresponding to $3\arcsec\times 3\arcsec$ at $z=6$). The color bar shows the mass-weighted stellar density values. White circles pinpoint the location of the six satellites, S1-S6, whose properties are summarised in Tab. \ref{tab:satmass}.
    \label{fig:alth_sat}
    }
\end{figure*}

\begin{figure*}
\centering
\includegraphics[width=0.98\textwidth]{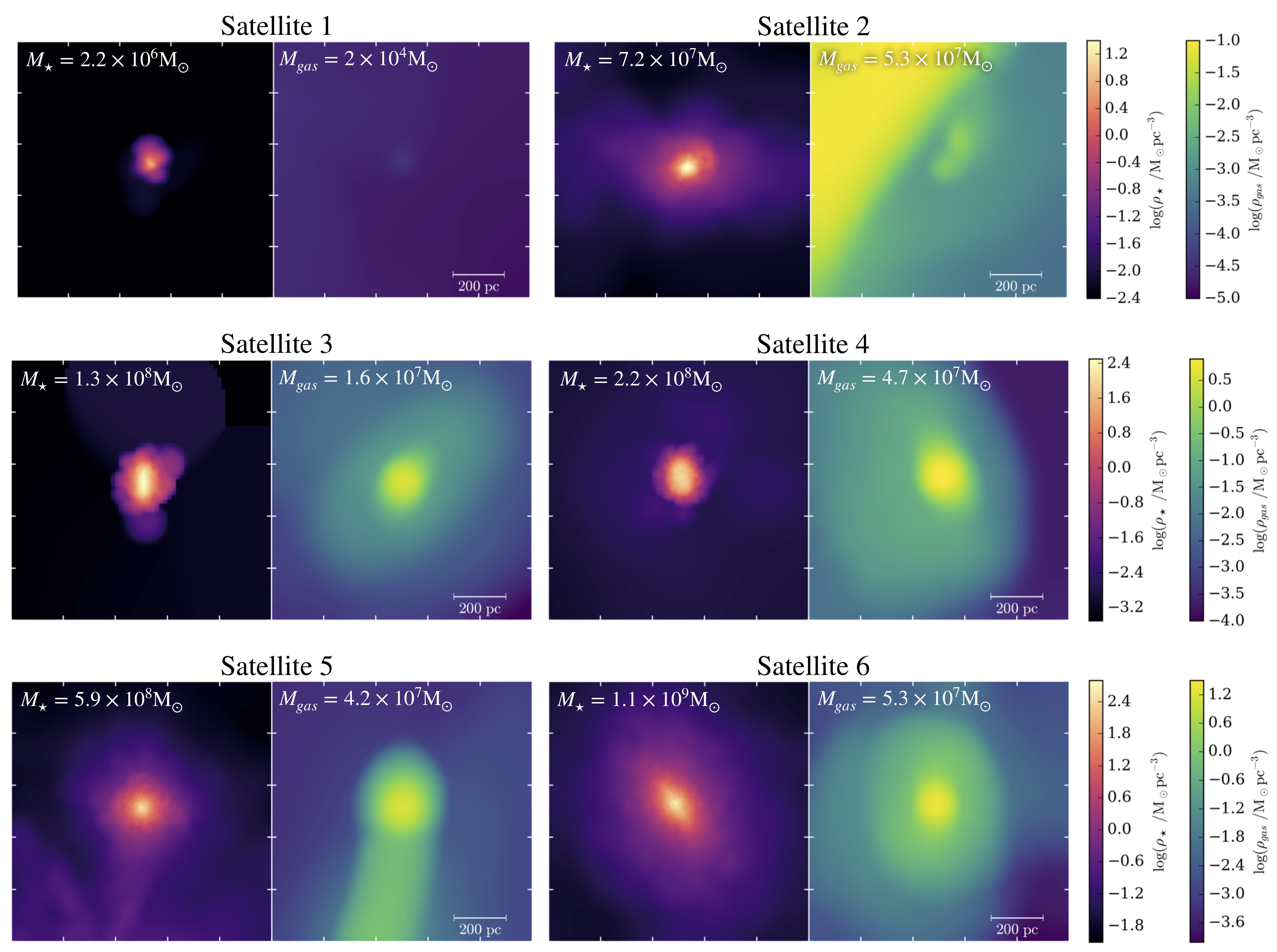}
\caption{Stellar and gas density maps of the six satellites. Each map is obtained by mass averaging the density along the line of sight in a box of size 1 kpc that is centred on the stellar centre of mass of the satellite.
\label{fig:mapdens}
}
\end{figure*}

\subsection{Identifying satellite galaxies}

We will hereafter focus our attention on dwarf galaxy satellites hosting stars, which appear in Fig.~\ref{fig:rad_althaea} as peaks in the stellar density profile (also identified by arrows).
In a forthcoming paper (Gelli et al. in prep.) we will study the peaks in the gas radial profiles 
(visible in the outer region, $\geq$ 8 kpc) that are instead associated to gas-rich, metal-poor, 
and {\it star-less} dwarf galaxy satellites.
In order to properly individuate dwarf satellites hosting stars within the inner regions of the 
virial volume of \althaea, we take advantage of stellar mass density maps like the one shown in Fig. \ref{fig:alth_sat}. We perform our search in the displayed region of 20 kpc $\times$ 20 kpc, 
chosen as to correspond at $z\simeq6$ with the field of view of the NIRSpec instrument of the JWST, i.e. $3\arcsec\times 3\arcsec$.
The criteria used to identify dwarf satellites consists in looking for maxima in the stellar density field, using a threshold of $\simeq 10^{-2}~\voldpc$. We perform the search in two 2D maps with line-of-sights perpendicular to each other, and combine the results to derive the coordinates in the 3D simulated space. The method identifies stellar overdensities but not necessarily galaxies, since it does not use any information on the dark matter content, which is checked \textit{a posteriori}.

We find a total of six satellites hosting stars (white circles of Fig. \ref{fig:alth_sat}). They are numbered from S1 to S6 from the least to the most massive in terms of stellar mass.
Tab. \ref{tab:satmass} reports their stellar ($M_\star$), gas ($M_{\rm gas}$) and halo ($M_{h}$) mass, along with the distance from the centre of the host galaxy, $D_{\rm \althaea}$. 
The stellar and gas masses are computed by summing the contribution within a box of size 1~kpc centred on the satellites. While such selected region is large enough to fully include the stellar component, this is not true for the DM component which is likely more extended. Furthermore we recall that the dwarf satellites are embedded in the environment of \althaea. For this reason we report $M_{h}$ through the cosmological baryon fraction as $M_{h}= (\Omega_m/\Omega_b) (M_\star+M_{gas})$, which represents a lower limit of the actual halo-mass.

The stellar masses of the six objects range from $2.2 \times 10^6 \msun$ of S1 to $1.1 \times 10^9 \msun$ of S6; the gas masses are about ten times lower, in the range $\simeq 10^{5-8} \msun$.
We will hereafter classify the satellites in three categories based on their stellar mass: low-mass (S1 and S2), intermediate (S3 and S4), and high-mass satellites (S5 and S6).
\begin{table}
  \begin{center}
    \caption{Main properties of \althaea's satellites found at $z=6$. The columns indicate (a) the satellite ID (indexed by increasing stellar mass), (b) distance from \althaea, (c) stellar mass, (d) gas mass, (e) halo mass. The stellar and gas masses are computed within a box of size 1~kpc centred on the satellite. The halo mass is computed as $M_{h}=(\Omega_m/\Omega_b)(M_\star+M_{gas})$.
    \label{tab:satmass}
    } 
    \begin{tabular}{|c|c|c|c|c|}
      \hline\hline
      Satellite & $D_{\rm \althaea}$ & $M_\star$      & $M_{gas}$         & $M_h$  \\
      ~         & $[\kpc]$           & $[10^8 \msun]$ & $[10^8 \msun]$    & $[10^8 \msun]$\\
      (a)       & (b)                & (c)            & (d)               & (e)           \\
      \hline
      S1        & 11.42              & 0.022          & 2$\times 10^{-4}$ & 0.16 \\
      S2        & 2.45               & 0.72           & 0.53              & 5.36 \\
      S3        & 6.85               & 1.3            & 0.16              & 8.95 \\
      S4        & 2.85               & 2.2            & 0.47              & 16.3 \\
      S5        & 1.23               & 5.9            & 0.42              & 40.3 \\
      S6        & 4.86               & 11.1           & 0.53              & 71.1 \\
      \hline\hline
    \end{tabular}
  \end{center}
\end{table}

\begin{figure*}
\centering
\includegraphics[width=0.9\textwidth]{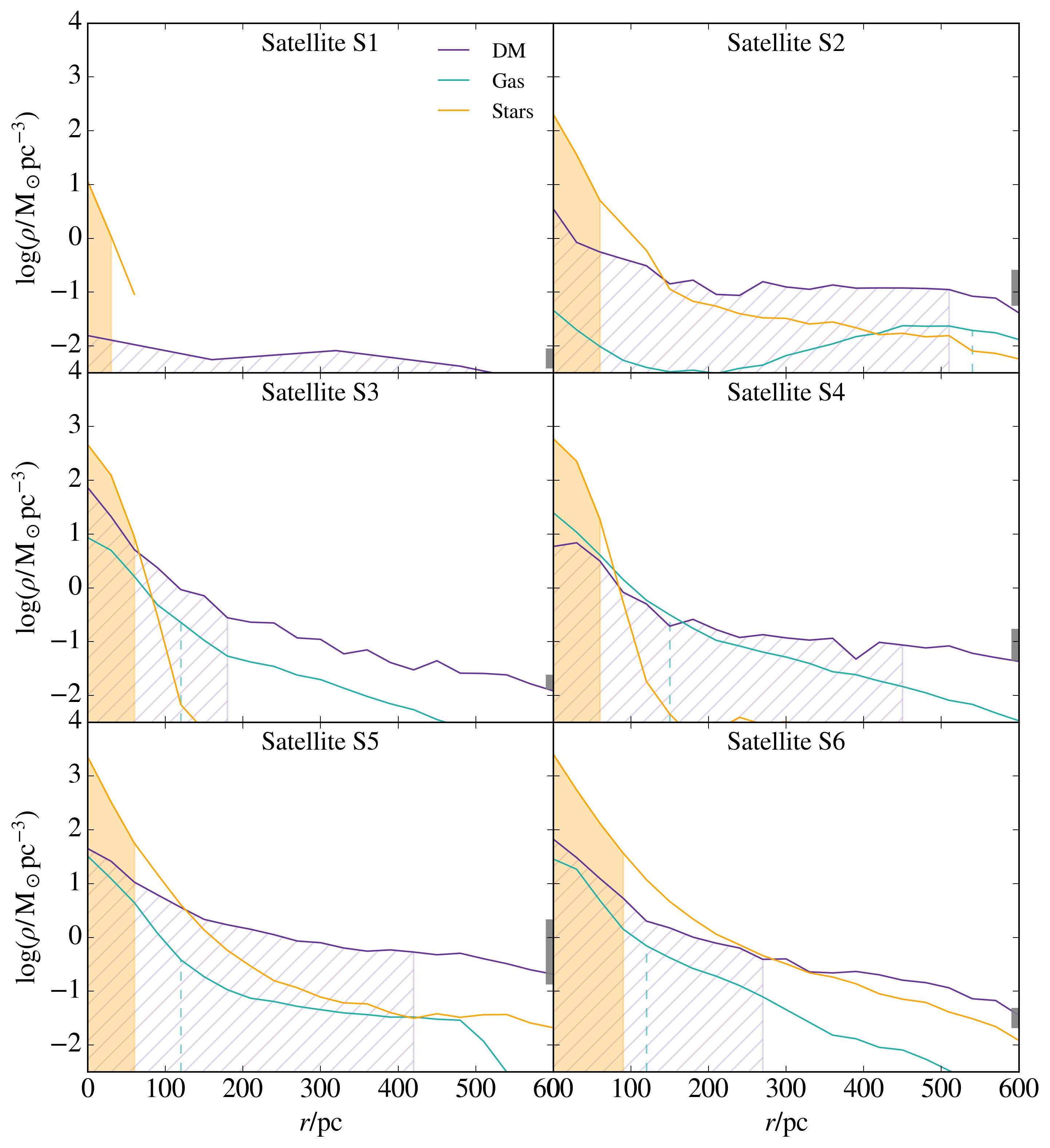}
\caption{Density profiles for the six satellite galaxies. For each satellite, we plot spherically-averaged mass densities for the stellar, gas, and dark matter component. The shaded orange area, the dashed green line and the grey hatched area indicate the regions enclosing 50\% of the stellar, gas and DM mass, respectively. The vertical thick line at $r= 600\, \rm pc$ shows the range of values of the DM profile of \althaea~(Fig. \ref{fig:rad_althaea}) between $r=D_{\rm \althaea}\pm600 \,\rm pc$. See Tab. \ref{tab:satmass} for the total masses of individual satellites.
\label{fig:rad_tot}
}
\end{figure*}

\begin{figure*}
\centering
\includegraphics[width=0.9\textwidth]{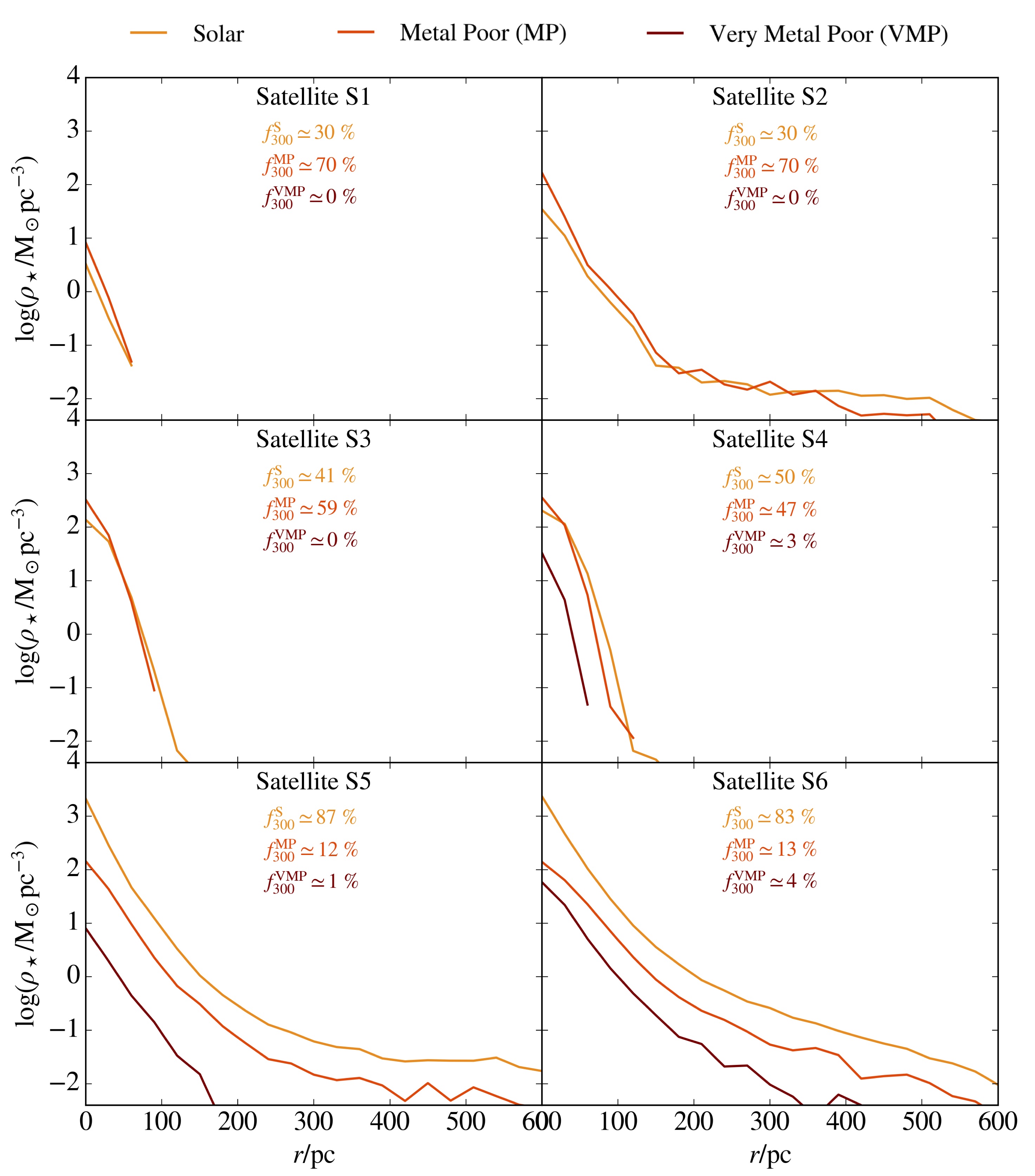}
\caption{Radial profiles of the different stellar populations in each satellite.
For each satellites we distinguish different populations based on three metallicity intervals: Solar $\log{Z_\star/\Zsun} \geq -0.5$, Metal Poor (MP) $-2 \leq \log{Z_\star/\Zsun} < -0.5$ and Very Metal Poor (VMP) $\log{Z_\star/\Zsun} < -2$. Similarly to Fig. \ref{fig:rad_tot}, profiles are spherically-averaged and mass weighted. We also indicate the mass fraction of Solar ($f^S_{300}$), MP ($f^{MP}_{300}$) and VMP ($f^{VMP}_{300}$) stars in the inner 300 pc of each satellite.
\label{fig:radial_z_bins}
}
\end{figure*}

\section{Dwarf satellites of high-$z$ LBGs}\label{sec_properties} 

\subsection{Density maps}\label{subsec_densmaps}

In Fig. \ref{fig:mapdens} the stellar/gas density maps the six satellites are shown. In low- and intermediate-mass satellites, stars are concentrated in the inner $\simeq 200\,\rm pc$, where $\rho_\star > 10^{-2} \voldpc$. More massive objects, i.e. S5 and S6, show a notable diffuse stellar component, with $10^{-2} < \rho_\star / \voldpc < 1$ in the outer $r > 200 \,\rm pc$ regions. In all satellites, however, the stellar density distribution peaks at the centre ($r < 50\,\rm pc$), is very close to spherical, and -- differently from the main galaxy \althaea\, -- exhibits no hints of a disk (see Fig. \ref{fig:alth_sat}).
In most cases the gas distribution tracks the stellar one in the central regions, but it then extends up to larger radii, sometimes filling the entire field of view ($1\, {\rm kpc} \times 1\, {\rm kpc})$. 

We note the peculiar case of the least massive and furthest satellite S1, whose density map indicates a considerable lack of gas. The little gas present ($M_{gas}=2\times 10^4 \msun$) has an extremely low density, $\rho_{\rm gas}\simeq 10^{-4} \voldpc$, and a very peculiar distribution.
Another quaint feature can be found in S2. In this satellite the gas is concentrated in two well-separated clumps, displaced from the bulk of the stellar population. S2 is located at the edge \althaea's disk (Fig.~\ref{fig:rad_althaea}, also seen as a density discontinuity in Fig. \ref{fig:mapdens}). It is unclear whether the ongoing interaction might be responsible for the displacement. However, this interpretation is possibly supported by the presence of two stellar streams in S2 which might result from tidal forces. Stellar streams are also found in S5, which is the closest to \althaea~($D_{\rm \althaea}=1.23 \,\rm kpc$), while they are not seen in more distant satellites.
We further discuss the nature of these streams in App. \ref{app:streams}.

\begin{figure*}
\centering
\includegraphics[width=0.96\textwidth]{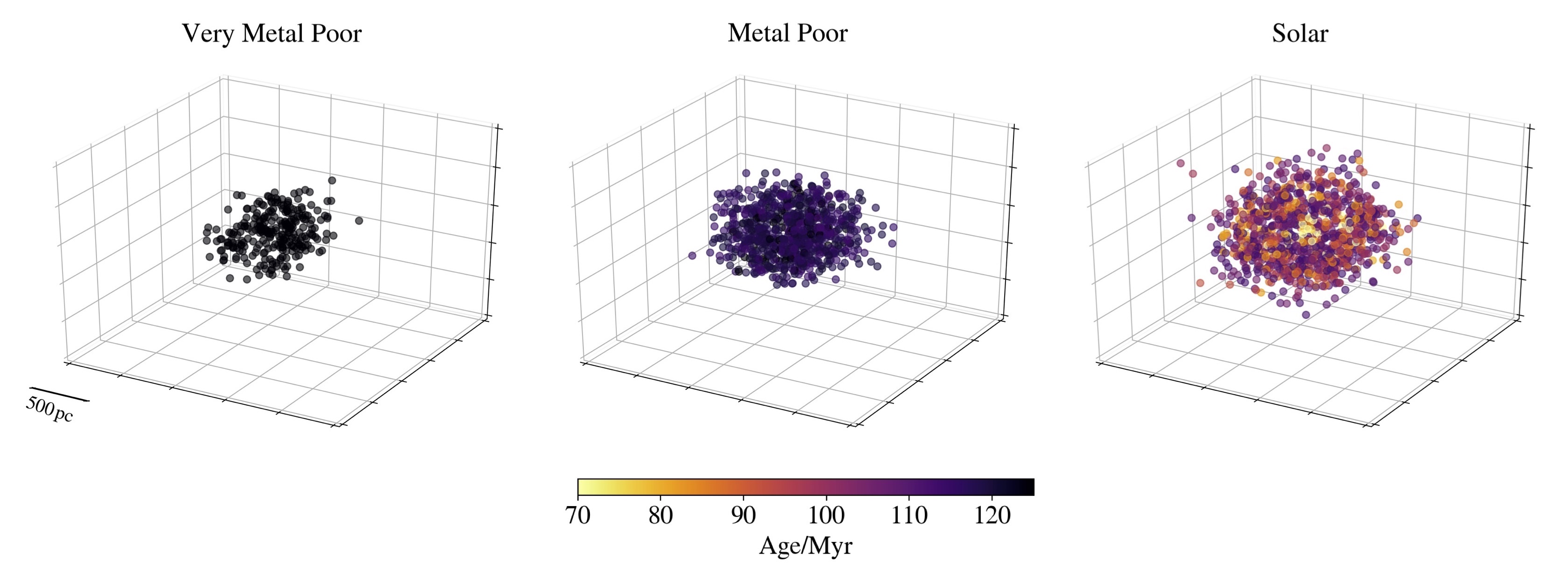}
\caption{3D spatial maps of an intermediate mass satellite S4 ($M_\star=2.2\times10^8\msun$) at $z=6$. Stars are divided in three metallicity intervals as in Fig. \ref{fig:radial_z_bins}. In the Metal-Poor and Solar ranges we show $1/6$ of the particles, selected randomly, in order to reduce superposition effects. The colorbar indicates the age of each stellar particle.
\label{fig:S4_3Dmaps}
}
\end{figure*}

\subsection{Radial profiles}\label{radialprofiles}

A more quantitative interpretation of the density field properties of gas, stars, and dark matter can be inferred through the spherically-averaged radial density profiles shown in Fig.~\ref{fig:rad_tot}. We immediately notice two common features among these systems: (a) stars dominate the total mass in the inner regions ($r<100 \,\rm pc$); (b) the central density increases with the satellite $M_\star$, from $\rho_\star\simeq1\,\voldpc$ for S1 ($M_\star = 2.2 \times10^{6}\msun$) to $\rho_\star\simeq10^3\voldpc$ for S6 ($M_\star = 1.1\times10^{9}\msun$).

The low- and intermediate-mass satellites have the entirety of their stars confined in the inner 100 pc. The only exception is represented by satellite S2, whose stellar density flattens after $\simeq 150\,\rm pc$ due to the presence of the stellar streams visible in the density map (Fig.~\ref{fig:mapdens}). S2 also has a peculiar gas distribution with an independent increase at high radii caused by the diffused gas surrounding \althaea~itself. 

The dark matter masses range between $5\times 10^6 \msun \lesssim M_{\rm DM} (r<600 \rm pc) \lesssim 5\times 10^8\msun$ in the displayed inner regions of the satellites. Its profile has typically a central peak as well, but it shows a more uniform distribution throughout the galaxy: it slowly decreases with radius until it reaches the underlying profile of the dark matter halo of \althaea. This is highlighted in each plot by the grey bat on the right axis indicating the DM density range in \althaea~(purple curve in Fig. \ref{fig:rad_althaea}) in $r=D_{\rm \althaea}\pm 600 \,\rm pc$.

In all satellites, the gas mass represents $\approx 10\%$ of the stellar mass. This is not unusual for dwarf galaxies since, due to their shallow potential well, they are unable to retain large amounts of gas which is easily evacuated.

We point out the interesting profile of the smallest satellite S1, characterised by the lowest values of stellar, gas, and dark matter densities. Due to the low amount of dark matter particles in the selected volume of $1\, \rm kpc^3$, in order to obtain a radial profile we averaged over a length of 150 pc, i.e. larger than the spatial resolution. The resulting profile shows the floor value of \althaea~dark matter halo, with no evidence of a central dark matter density peak.
As a consequence, S1 cannot be considered as a dwarf galaxy. However, the mean velocity of the simulated stellar particles in the satellite centre of mass reference frame is $\rm \langle v\rangle \simeq 14\,\kms$, lower than the escape velocity of $\rm v_{esc} \simeq 19 \,\kms$. This suggest that this object is gravitationally bound, and has virtually no dark matter: a situation akin to what recently measured in (proto-)globular clusters \citep[e.g.][]{Vanzella17}.

\subsection{Stellar metallicity}\label{stellarmetallicity}

In order to investigate the stellar populations of the satellites, we divided their stars into three categories according to their metallicity. We will hereafter refer to {\it Solar metallicity} stars if $\log{(Z_\star/\Zsun)} \geq -0.5$, to {\it Metal Poor} stars (MP) if $-2 \leq \log{(Z_\star/\Zsun)} < -0.5$, and to {\it Very Metal Poor} stars (VMP) if $\log{(Z_\star/\Zsun)} < -2$.
Fig. \ref{fig:radial_z_bins} shows the spherically-averaged density profiles of these three stellar populations. VMP stars are only present 
in the most massive satellites S4, S5, and S6 ($M_{\star}\gsim 1.5 \times 10^8 \msun$); they are always located in the inner regions, and constitute a subdominant stellar population (VMP mass fraction $< 5\%$ at all radii). 

MP stars are also a sub-dominant population in the most massive satellites (S5 and S6), which at all radii are dominated by Solar metallicity stars ($80-90\%$ of the total stellar mass). Interestingly, in intermediate- (S3 and S4) and low-mass (S1 and S2) satellites MP stars represent on average $60-70\%$ of the total stellar mass, and this fraction is even higher in the inner regions.

To understand how different stellar populations are spatially distributed, in Fig. \ref{fig:S4_3Dmaps} we show a rendering of the 3D positions of stellar particles in a typical dwarf satellite (the intermediate-mass S4), color coded with their age at $z\simeq6$. VMP stars are the oldest, they are more compact and more centrally-concentrated. Conversely, all very young stars have $Z\simeq \Zsun$ and they are more extended. In spite of these differences, all three stellar populations show a remarkably spherical distribution.

\begin{figure}
\centering
\includegraphics[height=0.98\textheight]{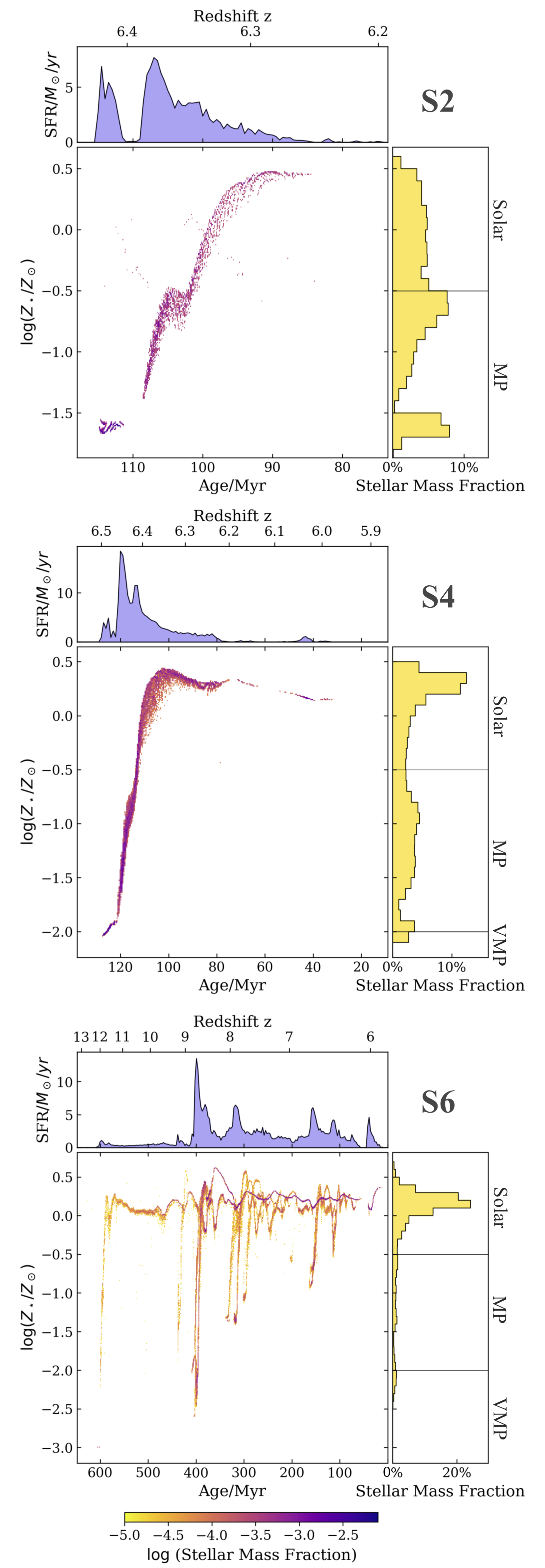}
\caption{(Caption in the next column.)
\label{fig:histories}
}
\end{figure}

\addtocounter{figure}{-1}
\begin{figure}
\caption{
From top to bottom, the three panels show the results for a low-mass (S2), an intermediate-mass (S4), and a high-mass (S6) satellites at $z\simeq6$. 
\textit{Main panel}: Metallicity of star particles inside a box of 1 kpc vs their age. The colorbar is the mass-weighted probability distribution function. 
\textit{Upper inset}: Star formation rate (SFR) as a function of stellar age (lower axis) and redshift $z$ (upper axis). 
\textit{Right inset}: Metallicity Distribution Function averaged in intervals of $\Delta(\log Z/Z_\odot)=0.1$.
}
\end{figure}

\section{Formation and evolution}\label{formationandevolution}

We now turn to the analysis of the star formation and metal enrichment histories of individual satellites. These are shown in Fig.~\ref{fig:histories} for three illustrative satellites, one for each of the different mass range categories: S2 (low-mass), S4 (intermediate-mass) and S6 (high-mass).
The main panel for each satellite shows the stellar age-metallicity relation, i.e. the metallicity of stellar particles as a function of their age. The colours represent the mass weighted probability distribution function (PDF) in this $Age-Z_\star$ plane.
The right and top insets show the integrated quantities for each satellite: the stellar metallicity distribution functions (MDF) at $z\simeq6$ and the star formation histories (SFH), i.e. the star-formation rate ${\rm SFR}(t) = dM_\star/dt \,[\msunyr]$ as a function of stellar age (or formation redshift). 
\subsection{Low-mass satellites}
In low-mass systems (S2), the SFH is remarkably simple and short, lasting $\simeq 30$ Myr only. All stars are formed in two bursts, separated by a short period of inactivity, and characterised by a SFR that reaches values of $7\,\msunyr$. This kind of intermittent SF behaviour represents a common feature of low- and intermediate-mass systems; it can be understood as follows.
The first burst of SF lasts for $\simeq 5$ Myr, after which the SF is inhibited. This is the typical timescale for the evolution of massive stars ($40\,\msun$) that end their lives as energetic SNe, strongly heating up the gas because of their high specific luminosity.\footnote{This is a consequence of the fact that the interstellar medium (ISM) where they act has already been pre-heated by stellar radiation and winds \citep{Agertz13}.}
The SF starts again after this transitory quenching phase of $\sim5$ Myr,
which is of the order of the cooling time computed in the typical conditions of the ISM after a SN event ($T\sim10^6$ K, $n\sim10^{-2}\rm cm^{-3}$ and $Z_{\rm ISM}\sim10^{-2}\zsun$). 
Afterward in few Myr the SFR reaches the second peak of $8\,\msunyr$. The rate is damped again after $\sim 5$ Myr, and then it continues to decrease until the SF is eventually quenched completely (after $\sim25$ Myr).

The $Age-Z_\star$ relation shows us that the first stellar generation formed in S2, at $z\sim6.4$, has high metallicity: $Z_{\star, in}\simeq 10^{-1.7} \zsun$. Dwarf satellites are indeed born from the gaseous environment surrounding the central LBG \althaea, which started forming stars at $z\sim15$ and since then has been polluting its surrounding medium with metals injected via SN-driven outflows.

After the quenching period, stars produced by the second burst in S2 have higher metallicity, as they are born from an ISM polluted by the previous population. The formation then proceeds in a more regular fashion, and stars forming at later times are progressively more metal-rich, rapidly reaching super-solar values. We also notice a brief and slight decrease of $Z_\star$ (age $\simeq 105$ Myr) caused by metal loss via SN-driven winds. 

The stellar MDF of S2 at $z\simeq6$ reflects its SF and chemical enrichment history. It shows two maxima in correspondence of the two main SFR peaks; the fraction of MP stars is $65\%$ of the total.

\subsection{Intermediate-mass satellites}
The S4 history overall resembles that of S2, but a few differences are worth noticing: (i) the SF activity in S4 starts earlier ($z\sim6.5$) and hence its first stellar population is slightly less enriched ($Z_{\star, in}\simeq10^{-2}\zsun$), (ii) the overall SFH lasts longer ($\sim 100$ \myr) and, (iii) the SFR reaches a higher peak value ($\sim15\,\msunyr$). 
As a consequence the MDF is rather evenly distributed in the metallicity range $10^{-2}\zsun-10^{0.5}\zsun$ but the peak appears at higher metallicities, i.e. in correspondence with the longer, final SF phase. Thus, the fraction of MP stars is somewhat lower, now representing $47\%$ of the total.

\begin{figure*}
\centering
\includegraphics[width=0.95\textwidth]{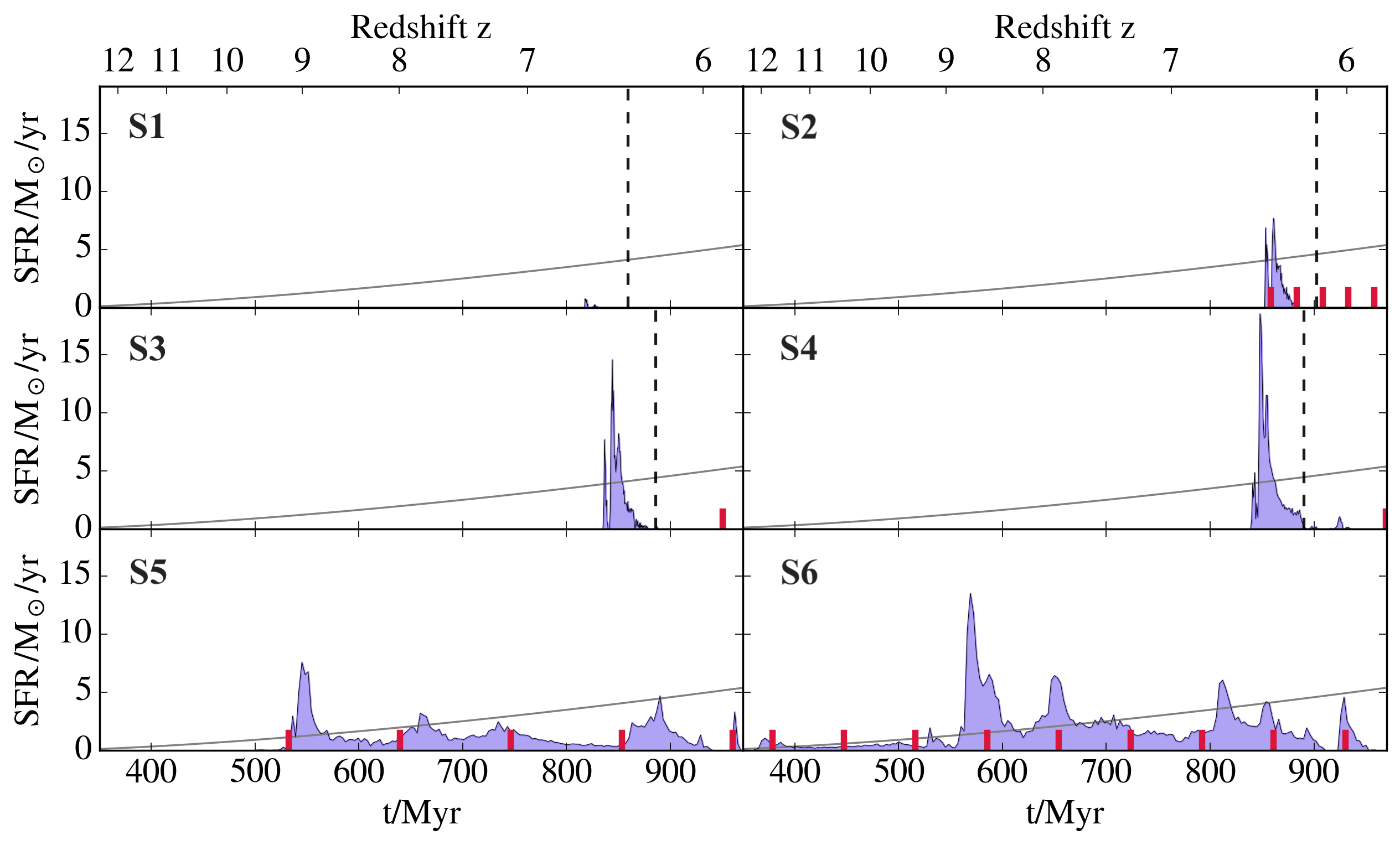}
\caption{Star formation histories of the six dwarf satellites as a function of the cosmic time (lower axis) and redshift (upper axis).
Red segments indicate the time at which the satellites were at their orbital periapsis; see Sec. \ref{orbits} and Tab. \ref{table:orbits} for details. The value of the interstellar radiation field, scaled for display purposes, $G(t)/20 G_0$, is overplotted as a grey line. For S1,S2,S3 and S4, the vertical dashed line indicates the last SN explosions associated to the peak of star formation.
\label{fig:All_sat_hist} }
\end{figure*}

\subsection{High-mass satellites}

The high-mass satellite, S6, has the most complex SF/metal enrichment history (Fig.~\ref{fig:histories}). Its first stellar generations are formed at $z\simeq 13$; afterwards the satellite undergoes a continuous SF activity up to $z\simeq6$, i.e. for $\approx 600$ Myr. Contrary to lower-mass satellites, the initial phases of SF in S6 (and S5) are characterised by a moderate $SFR \approx 1-2\msunyr$, which suddenly increases at $z\approx 9$ with a first and highest peak of SFR $\approx 15\msunyr$, followed by many other weaker bursts, with the latest one occurring at $z\approx 6$.

The $Age - Z_\star$ relation shows that the first stellar generations, formed at $z\approx 13$, have the lowest metallicity, $Z_\star = Z_{floor} = 10^{-3}\zsun$, meaning that their formation happened in an almost pristine\footnote{We remind that, as we cannot resolve mini-halos, we have set an artificial metallicity floor to mimic their enrichment. Thus, we cannot exclude that the first stellar generation in S6 is indeed made of metal-free, PopIII stars.} chemical environment. However, following the first SF events, the enrichment proceeds rapidly, and the ISM metallicity reaches solar values in $<100$ Myr. Interestingly, the $Age - Z_\star$ plane is characterised by many vertical tracks. They indicate that stars with very low metallicities are continuing to form in the galaxy in spite of the global, average solar metallicity of the satellite. This can be attributed to two types of events: (i) infalling, quasi-pristine gas inducing a new burst of star formation, or (ii) acquisition of low-$Z$ stars due to a dry merger with a smaller, older system. In the latter case, a $Z_\star-Age$ relation similar to S2 and S4 should be found for the infalling stars, i.e. a short first burst followed by a longer and more intense one. Some of the vertical tracks in S6 indeed show this behaviour, but others do not. We are led to conclude that both processes are acting at the same time, as expected from the theoretical understanding of galaxy assembling through cosmic times.

The resulting MDF shows that S6 is the only satellite hosting VMP stars at the metallicity floor (and perhaps PopIII stars). These VMP stars only represent $\simeq4\%$ of S6 total stellar population, which is therefore largely dominated by Solar metallicity stars.

\section{What quenches star formation?}\label{sec_interpretation}
We next intend to isolate the physical processes driving the SFH of the satellites,ultimately determining their properties at $z\simeq6$. In Fig. \ref{fig:All_sat_hist} the SF/metal enrichment histories for all the six satellites are plotted as a function of cosmic time, $t$, in the range 350-950 Myr.

It is apparent that low- and intermediate- mass satellites (S1-S4) form all of their stars in two short bursts ($\lesssim50$ Myr) and are afterwards  completely quenched, while the high-mass ones (S5-S6) have remarkably longer SFHs, lasting $400-500$ Myr, i.e. up to $z\simeq6$.

We have already noted how the central LBG affects the chemical properties of stellar populations in the satellites. Those forming at later times have higher initial metallicities, $Z_{\star,in}$, because of the (pre-)enrichment of their birth environment, i.e. the LBG circum-galactic medium, by \althaea~outflows.
Does the presence of the central LBG also leave an imprint on the SFHs of the smallest satellites via, e.g. dynamical interactions or mechanical and/or radiative feedback induced by massive stars? In the following we review these two possibilities in detail.

\subsection{Quenching by dynamical interactions}\label{orbits}

\begin{figure*}
\centering
\includegraphics[width=0.66\textwidth]{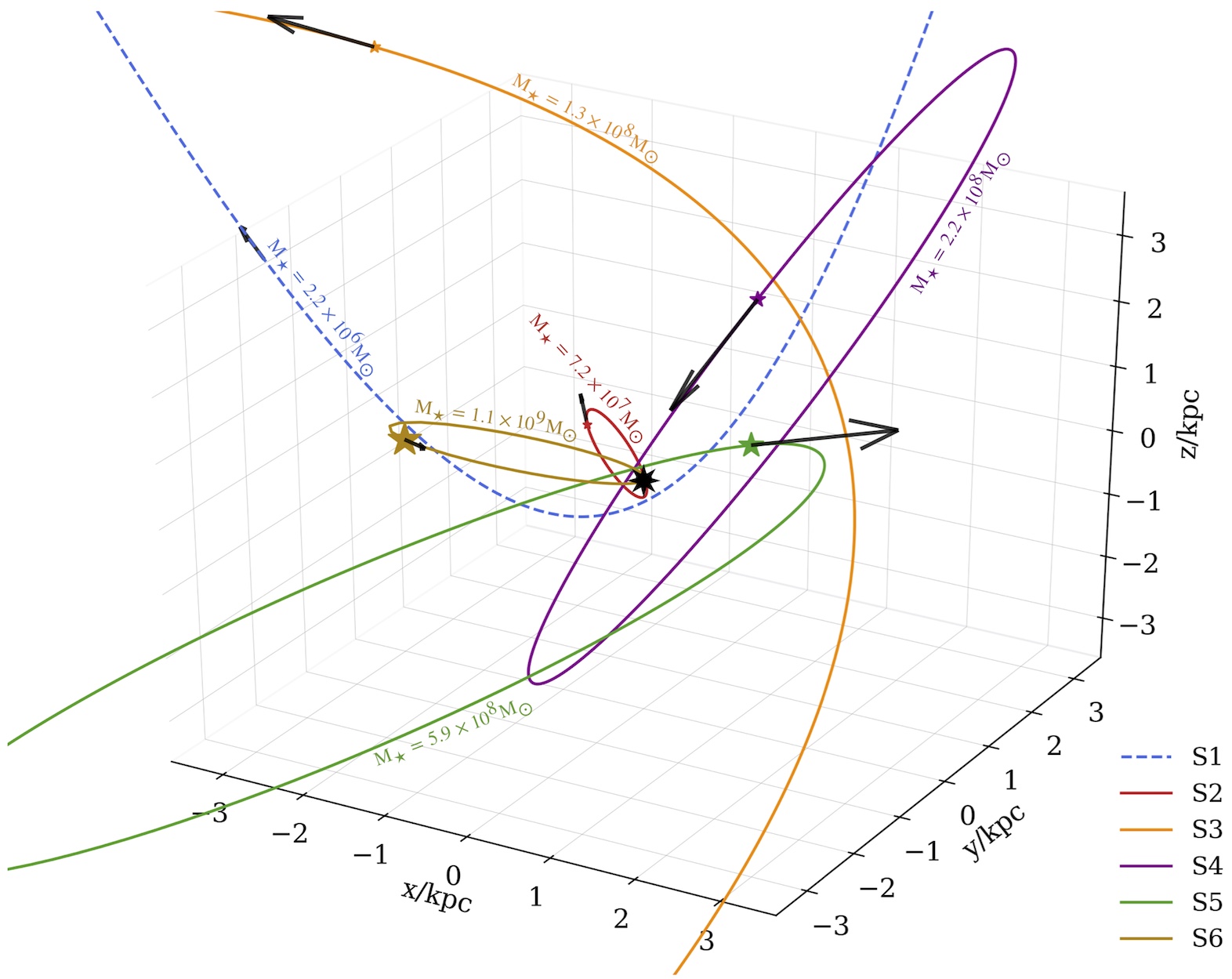}
\caption{Orbits of the six dwarf satellites around the central galaxy \althaea~(central black star). The stars and arrows indicate the location and the velocity direction of the satellites at $z\simeq6$, and their sizes are proportional to the stellar mass and to the velocity respectively. The dashed blue line correspond to the open hyperbolic trajectory of satellite S1. 
\label{fig:orbit}
}
\end{figure*}
The orbital motion of the dwarf satellites is mainly determined by the gravitational influence of the central massive galaxy \althaea. While forming stars, a satellite may pass through regions of varying density and even near the central galaxy, possibly interacting with it. In order to shed light on the influence that such dynamical interactions have on the satellite properties, we compute the orbits of the satellites adopting a simple approach, similar to those applied to MW satellites \citep[e.g.][]{Simon19}, and which can provide a good approximation of their effective motions. 

We start by selecting a reference frame in which \althaea~is at rest at the origin, and assume that each satellite moves on an elliptical orbit around the central galaxy. Approximating the satellites as point-like masses, the orbital equations for the two-body problem\footnote{This approach neglects mutual interactions with other satellites.}:
\be
\mathbf{\ddot{r}} + \frac{G M_{\rm A}}{r^3} \mathbf{r} = 0\,,
\ee
where $M_{\rm A}$ is the mass enclosed within a sphere of radius $D_{\rm \althaea}$ (i.e. the distance of the satellite from the central galaxy at $z\sim6$) centred on \althaea. Working in the plane identified by the satellite velocity vector, $\mathbf{v}_{z\simeq6}$, and the position of \althaea, and further using polar coordinates $r-\theta$, we impose two initial conditions: $r(\theta=0)= D_{\rm \althaea}$ and $\mathbf{v}(\theta=0)=\rm \mathbf{v}_{z\simeq6}$. Then, 

\be
r(\theta) = \frac{D_{\rm \althaea}}{(1-\beta^2)\cos\theta - ({v_r}/{v_t})\sin\theta + \beta^2};
\ee
$v_r$, $v_t$ are the radial and tangential components of $\mathbf{v}_{z\simeq6}$, respectively; $\beta= v_c/v_t$ involves \althaea's circular velocity, $v_c = (GM_A/D_{\rm \althaea})^{1/2}$. In Tab. \ref{table:orbits} we show the main parameters describing the orbits found: the radius at the periapsis $r_{\rm min}$; the time difference between the current instant ($z\simeq6$) and the time at which the satellite was at the periapsis $\rm \Delta T^{\rm near}$; the orbital period T; the eccentricity $e$; the modulus of the bulk velocity of the satellite $|\textbf{v}_{z\simeq6}|$; and the escape velocity $v_{\rm esc}$ from \althaea~at the satellite initial location.

In Fig.~\ref{fig:orbit} we show the orbits derived for the six dwarf satellites. The least massive satellite, S1, has no closed orbit solution and its bulk velocity (187 km/s) exceeds the local escape velocity (160 km/s). We conclude that not only S1 is not a proper galaxy -- given the absence of a dark matter overdensity (see Sec. \ref{radialprofiles}) -- but it is also not gravitationally bound to the halo in which it is located.

The velocities of the other satellites with respect to \althaea~vary in the range $\sim(60- 360)$ km/s; two of them (S2 and S6) have highly eccentric orbits ($e>0.9$) and small periapsis ($r_{\rm min}<1$ kpc). The orbital periods range from $T=25$ Myr to $T>\rm t_{ Hubble}$ like in the case of S3, meaning that this dwarf galaxy has not yet completed an orbit at $z=6$. Despite this, all satellites have $\rm \Delta T^{ near} < 70$~Myr, which implies that they all crossed the periapsis within the last 70 Myr (see Tab. \ref{table:orbits}).

The time of the periapsis position is a key quantity to interpret the SFHs since at the periapsis the satellites passed across the innermost and densest regions of \althaea~and thus it is most likely for the satellite to have directly interacted with it, for example through gas accretion or tidal effects. To interpret the SFHs, we here assume that the dwarf satellites have always moved along the orbits shown in Fig.~\ref{fig:orbit}. Yet, we should stress that such estimate works successfully only over brief periods of time ($\simeq {\rm few}\times$10 Myr around $z\sim6$) since on longer timescales the mass of the two bodies (i.e. the satellite and \althaea) might change significantly due to merger processes; multi-body interactions can also become important.

With this caveat in mind, we can compare in Fig. \ref{fig:All_sat_hist} the time of the periapsis position of each satellite (red vertical lines) with their SFHs. For low- and intermediate-mass satellites (S2, S3, S4) no connection can be found between the orbital motion and the shape of the SFH. In particular, these small satellites stopped forming stars hundreds of Myr before passing at the periapsis. For the two high-mass satellites (S5 and S6) instead, we see that the periapsis positions roughly correspond to the peaks of the SFR. This suggests that these satellites likely experienced compression and acquired new gas while passing through \althaea's densest and innermost regions.  

We conclude that dynamical interactions around the central LBG \althaea~do not represent the key process quenching the SF in the smallest satellites, although they may lead to non-negligible tidal effects, as discussed in App. \ref{app:streams}. Dynamical interactions play instead a key role in regulating and timing the SF activity of the largest satellites.

\begin{table}
  \begin{center}
     \caption{Orbital elements of each satellite: distance of the periapsis ($r_{\rm min}$), time of the periapsis ($\rm{ \Delta T^{ near} }= t_{z\simeq6} - t_{r_{\rm min}}$), period of the orbit (T), eccentricity ($e$), modulus of the velocity ($v_{z\simeq6}$), and 
     escape velocity ($v_{\rm esc}$).
    See the text in Sec. \ref{orbits} for the details of the calculation.
    \label{table:orbits}
    } 
    \begin{tabular}{|c|c|c|c|c|c|c|}
      \hline\hline
      Element & S1 & S2 & S3 & S4 & S5 & S6  \\
      \hline
      $r_{\rm min}[\kpc]$ &  --  & 0.1 & 2.3  & 1.9  & 1.0 & 0.02 \\
      \hline
      $\rm \Delta T^{ near} [\myr]$ &   --  & 12 &  20 &  0 & 10 & 69 \\
      \hline
     T [\myr] &  --   & 25 & 3990 &  505 & 107 & 40 \\
      \hline
     $e$ &  --   & 0.92 & 0.92  & 0.84  & 0.69 & 0.99 \\
      \hline
      $v_{z\simeq6} [\rm km\, s^{-1}]$ & 187  & 63 & 199 & 290  & 360 &  80\\
      \hline
      $v_{esc} [\rm km\, s^{-1}]$ & 160 & 320 & 208 & 314 & 247 &  247\\
      \hline\hline
    \end{tabular}
  \end{center}
\end{table}

\subsection{Quenching by feedback processes}\label{feedback}

Feedback processes are a key factor influencing the evolution of dwarf galaxies. In particular, we can distinguish between two negative feedback effects: (i) mechanical feedback, i.e. SN explosions occurring in the dwarf galaxy heating/removing the gas; (ii) external radiative feedback linked to the presence of the nearby LBG, which produces photons that dissociate \HH~molecules preventing SF in the dwarf.
Mechanical feedback starts $\simeq 5$ Myr after the first SF episode, i.e. as soon as $M\simeq 40\,\msun$ stars end their lives as SNe. It continues for $\simeq 40$ Myr after the last SF event, i.e. until the explosion of $M\simeq 8\,\msun$ stars (see also Sec.~\ref{formationandevolution}). The second effect is instead linked to the SF activity in the main LBG galaxy;  massive star there produce  large amounts of Lyman-Werner band ($11.2<h\nu/{\rm eV}<13.6$) radiation capable of photo-dissociating \HH\, molecules via the so-called two-step Solomon process.

In our simulation the average flux of the UV interstellar radiation field (ISRF) in the Habing band ($6.0-13.6\,\rm eV$; \citet[][]{Habing68}), is evaluated as $G(t)=G_0({\rm SFR}(t)/ \msunyr)$, where the SFR is computed by considering all stars within the virial volume of \althaea, and $G_{0}=1.6\times 10^{-3} {\rm erg}\,{\rm cm}^{-2}\,{\rm s}^{-1}$ is the MW value\footnote{For the MW we can roughly assume $\rm{SFR}\approx 1\msunyr$.}. High $G$ values are expected for typical LBGs, as also inferred by observations of high-z LBGs \citep{Carniani17} and of local dwarf galaxies, which are often considered as their present-day counterparts \citep{Cormier15}.
The ISRF is largely dominated by \althaea, and assumed to be spatially uniform. Thus, at any time, all dwarf satellites are illuminated by the same UV flux independently of their location (Sec. \ref{sec_simulation} and \citealt{Pallottini17}).
This is a fair approximation in the proximity of the central LBG and hence for the six satellites which are all found within 12 kpc from it \citep{Behrens18,Pallottini19}.
\HH~self-shielding is taken into account in the simulation on a cell by cell basis through the \citet{Richings14} prescription, being expressed as a function of the \HH~column density, turbulence, and temperature.

For each satellite in Fig.~\ref{fig:All_sat_hist} we show both the time of the last SN explosion associated to the peak of SF, and the evolution of the impinging ISRF using the SFR of \althaea, which follows the simple law ${\rm SFR}(t)=1.5\,\log(t_\star/30\,\myr)$, with $t_\star$ the time since its first SF event (see Fig.~2 of \citealt{Pallottini17}). Hence, the ISRF steadily increases with time, reaching $G=100\, G_0$ at $z\approx 6$.

\begin{figure}
\centering
\includegraphics[width=0.47\textwidth]{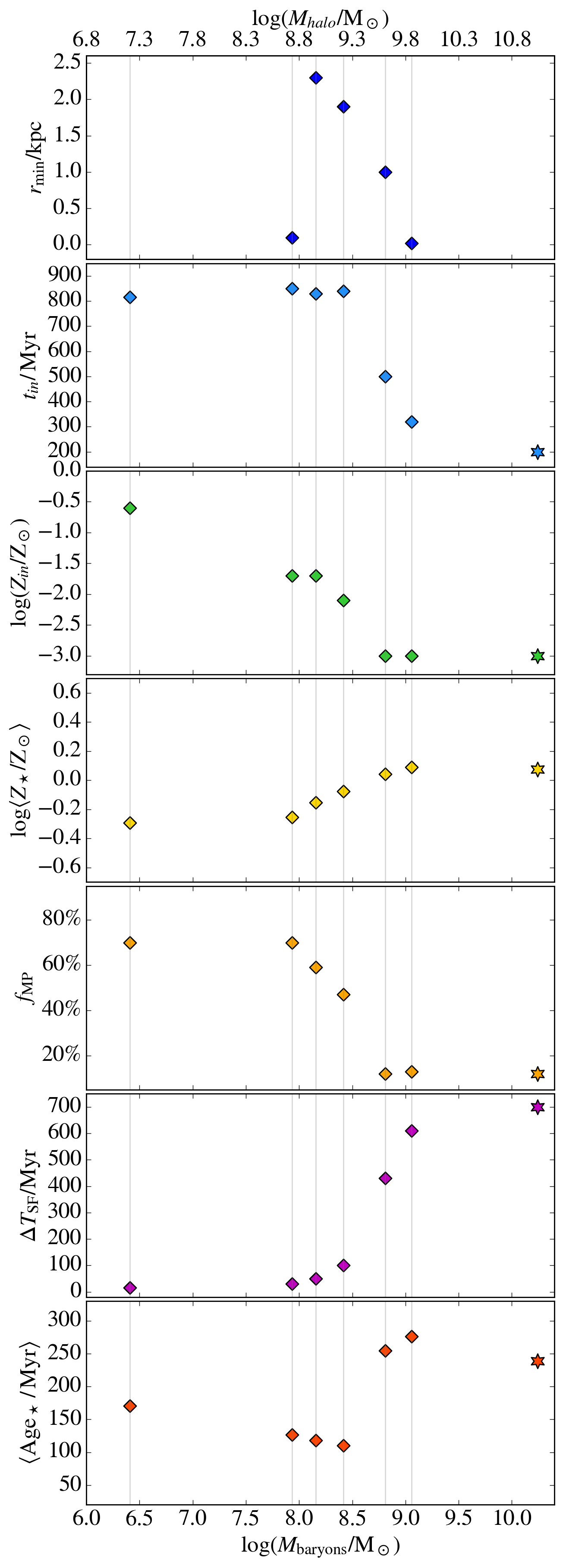}
\caption{(Caption in the next column.)\label{fig:prop_vs_mass}
}
\end{figure}

\addtocounter{figure}{-1}
\begin{figure}
\caption{Stellar properties of the six analysed dwarf satellites as a function of their stellar mass (lower axis) and sub-halo mass (upper axis). The star indicates the values for \althaea.
From top to bottom we plot: the periapsis potions of the satellites orbits ($r_{\rm min}$); the age of the Universe when the satellite formed ($t_{in}$); 
the metallicity of the oldest stellar population in the satellite ($Z_{in}$); the mass-weighted mean stellar metallicity at $z\simeq6$ ($\langle Z_\star \rangle$); the mass fraction of metal poor stars ($f_{\rm MP}$); the duration of the SFH ($\Delta T_{\rm SF}$); the mass-weighted mean stellar age $\langle \rm Age_\star \rangle$. 
}
\end{figure}

We are now ready to assess the impact of feedback on the SFHs of the satellites. We start our analysis from low- and intermediate-mass satellites. 
Fig.~\ref{fig:All_sat_hist} shows that their SF activity is concentrated between $t\sim800 - 900$ Myr and then they are completely quenched up to $z\simeq6$. The first decrease and then complete suppression of SF is caused by the energetic SN feedback that heats and ejects the gas.
The effect of radiative feedback is to dissociate \HH~molecules in the gas, but its contribution is not appreciable due to the low amount of remaining gas after SN explosions happened. We also notice that the ISRF does not vary significantly in the last $\sim200$ Myr and therefore has no crucial impact in changing the ISM conditions in such time interval.
From Fig.~\ref{fig:All_sat_hist} we see that in all these small systems the last SNe explosions happened $\geq 50$ Myr before, so that the gas might have had all the time to re-collapse and form stars (see Sec.~\ref{formationandevolution}). Despite this the satellites remain passive because: due to their small masses, and hence low binding energy, they are not able to pull new gas into their potential wells -- even when transiting through the inner regions of \althaea~-- and are thus unable to trigger SF.

The high-mass satellites S5 and S6 are also not forming stars at $z\simeq6$, but were active in the previous 5-10 Myr. This recent short period of inactivity is caused by internal SN explosions which are still ongoing at $z\simeq6$. In general these massive dwarf galaxies have long and uninterrupted SFHs for two reasons: (i) they better withstand SN explosions, i.e. the feedback is not as efficient in removing gas as in small satellites; (ii) they are massive enough to acquire fresh gas when passing near \althaea~(as discussed in Sec.~\ref{orbits}).

We conclude that mechanical feedback can quench SF in dwarfs around \highz LBGs, while radiative feedback has not a significant impact. Nevertheless the low amount of \HH~masses predicted in the simulated satellites ($ 0 < M_{\rm H_2}/\msun < 8\times10^4 $), are consistent with measurements from the Dwarf Galaxy Survey \citep{madden:2013,cormier:2014}, and models of low-$z$ dwarfs \citep[][]{lupi:2020}. However, because of the current limitations related to the radiative transfer prescriptions, a more robust investigation of this point is left for future work.

\section{Global properties}\label{sec_discussion}

Our analysis indicates that many of the observed properties of dwarf satellites are related to their stellar (hence, halo) mass. The halo mass sets the potential well and therefore determines the ability of the galaxy to retain its gas and survive dynamical interactions.
In Fig. \ref{fig:prop_vs_mass} we summarise the main properties of the dwarf satellites stellar populations as a function of baryonic mass (lower axis) and halo mass (upper axis): the periapsis potions of the satellites orbits $r_{\rm min}$; the formation time, $t_{in}$, and metallicity, $Z_{in}$, of their first stellar generation; the mass-weighted mean stellar metallicity at $z\simeq6$, $\langle Z_\star \rangle$; the fraction of metal poor (MP) stars present, $f_{\rm MP}$; the duration of the SF activity, $\Delta T_{SF}$; and the mass-weighted mean stellar age $\langle\rm Age_\star \rangle$.

From the top panel we see that high-mass satellites have lowest $r_{\rm min}$, allowing them to acquire mass while orbiting thorough the densest inner regions of \althaea. Also S2 have a small periapsis of 0.1 kpc: this near interaction with the central disk manifests with the creation of two stellar streams (see App. \ref{app:streams}).

Star formation activity starts relatively late for low- and intermediate-mass satellites ($800\lsim t_{in}/{\rm Myr} \lsim 850$); instead, the two most massive dwarfs formed the first stars much earlier (300 Myr after the Big Bang for S6). Consistently with this picture $t_{in}=200$~Myr for \althaea. As a consequence the initial metallicity (i.e. at formation) of the galaxies decreases with mass (third panel). Low- and intermediate-mass satellites have $Z_{in}=[10^{-2.5}-10^{-0.7}]\zsun> Z_{\rm floor}$, while the most massive S5 and S6 have $Z_{in}=Z_{\rm floor}=10^{-3}\zsun$ like \althaea.

Even though only the most massive dwarfs contain a VMP stellar component, they also have the highest mean stellar metallicity (fourth panel). While all systems are dominated by metal rich stars $\langle Z_\star \rangle > 0.3\,\zsun$, low-mass satellites contain a higher fraction of MP stars (fifth panel), similar to what observed in the Local Group \citep{Kirby13}. 

The duration of the SF activity $\Delta T_{SF}$ (sixth panel) strongly depends on stellar mass. Lower mass satellites have short SFHs and form stars is a few, short bursts without experiencing any merger events (see Sec. \ref{formationandevolution}). Higher mass systems are instead active for much longer periods during which fresh fuel is also brought in by merger episodes, allowing SF to proceed uninterruptedly for over 400 Myr.

As low- and intermediate- mass satellites formed at similar times $t_{in}$, the increase of $\Delta T_{SF}$ parallels the decreasing trend of $\langle \rm Age_\star \rangle = 100-150$ Myr (bottom panel); high-mass satellites present instead older populations, i.e. $\simeq 250$ Myr. Note that -- despite the earlier onset of the SF process -- the stellar population of~\althaea~is slightly younger than 250 Myr as a result of its increasing SFH (see Fig. \ref{fig:althaea_sfh}). Overall, the $\langle \rm Age_\star \rangle$ trend is broadly in agreement with that found by other theoretical works \citep[][in particular see Fig. 10 therein]{tacchella:2018}.

In summary, our findings are consistent with a scenario in which high-mass systems are the oldest and form out of an virtually unpolluted gas. Still, because of their high-mass and complex assembly history, they continue to form stars for a longer period of time, and hence are more metal-rich and host a smaller fraction of metal-poor stars. On the other hand, low-mass satellites are the youngest systems and form in environments that have been already pre-enriched by SN-driven outflows exploding in the central LBG. Yet, since they form stars for a short period of time, their metallicity is lower and thus they host a larger fraction of metal-poor stars.
Such behaviour is similar to that found by statistical models that study Local Group dwarf galaxies in a cosmological context \citep{Salvadori15} implying that this is a general outcome of $\Lambda$CDM hierachical structure formation.

\subsection{Comparing with \althaea}\label{sec_althaea}

\begin{figure}
\centering
\includegraphics[width=0.48\textwidth]{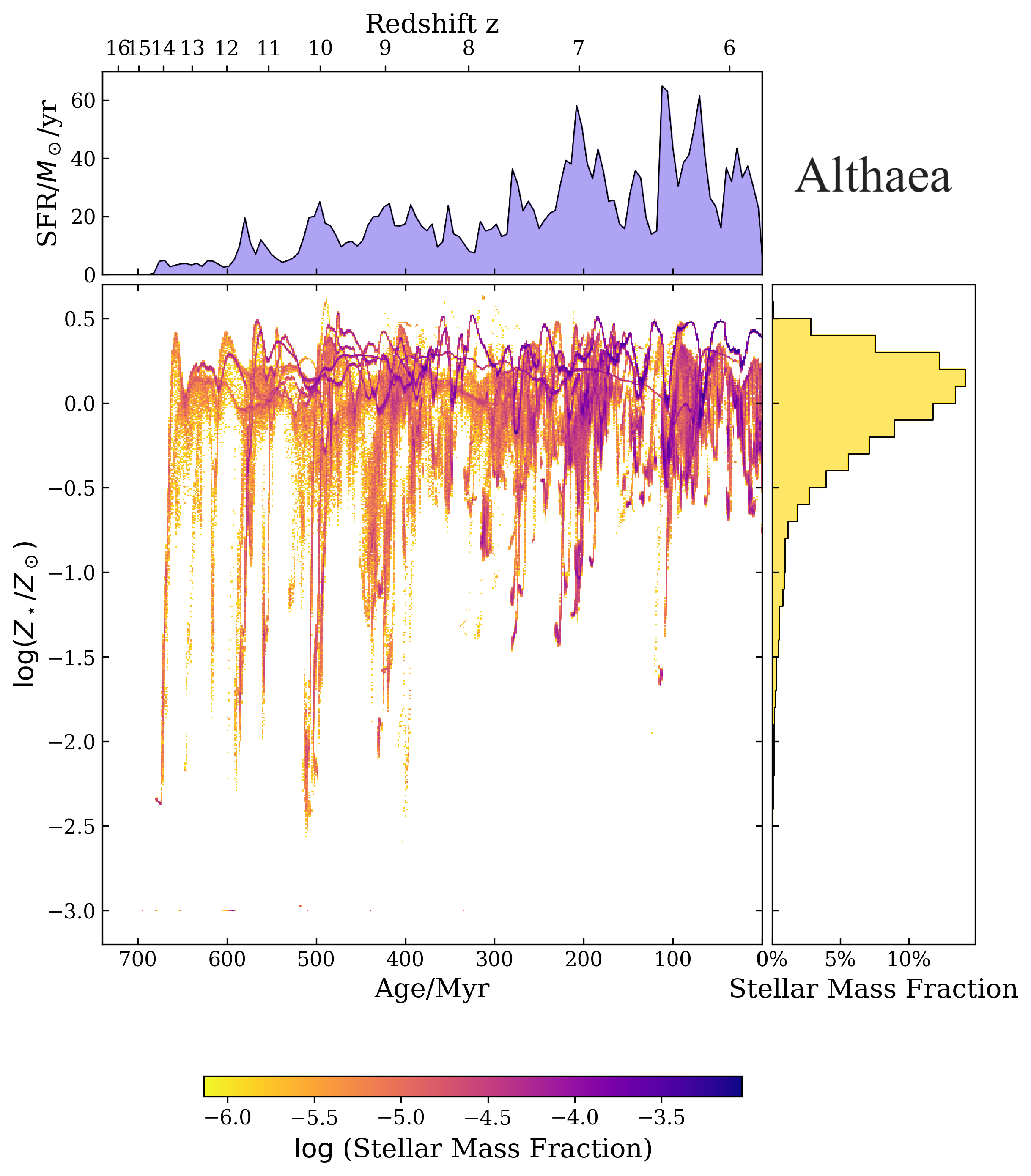}
\caption{ $Age-Z_\star$, Star Formation History and Metallicity Distribution Function of \althaea, as in Fig. \ref{fig:histories}.
\label{fig:althaea_sfh} }
\end{figure}

\begin{figure}
\centering
\includegraphics[width=0.48\textwidth]{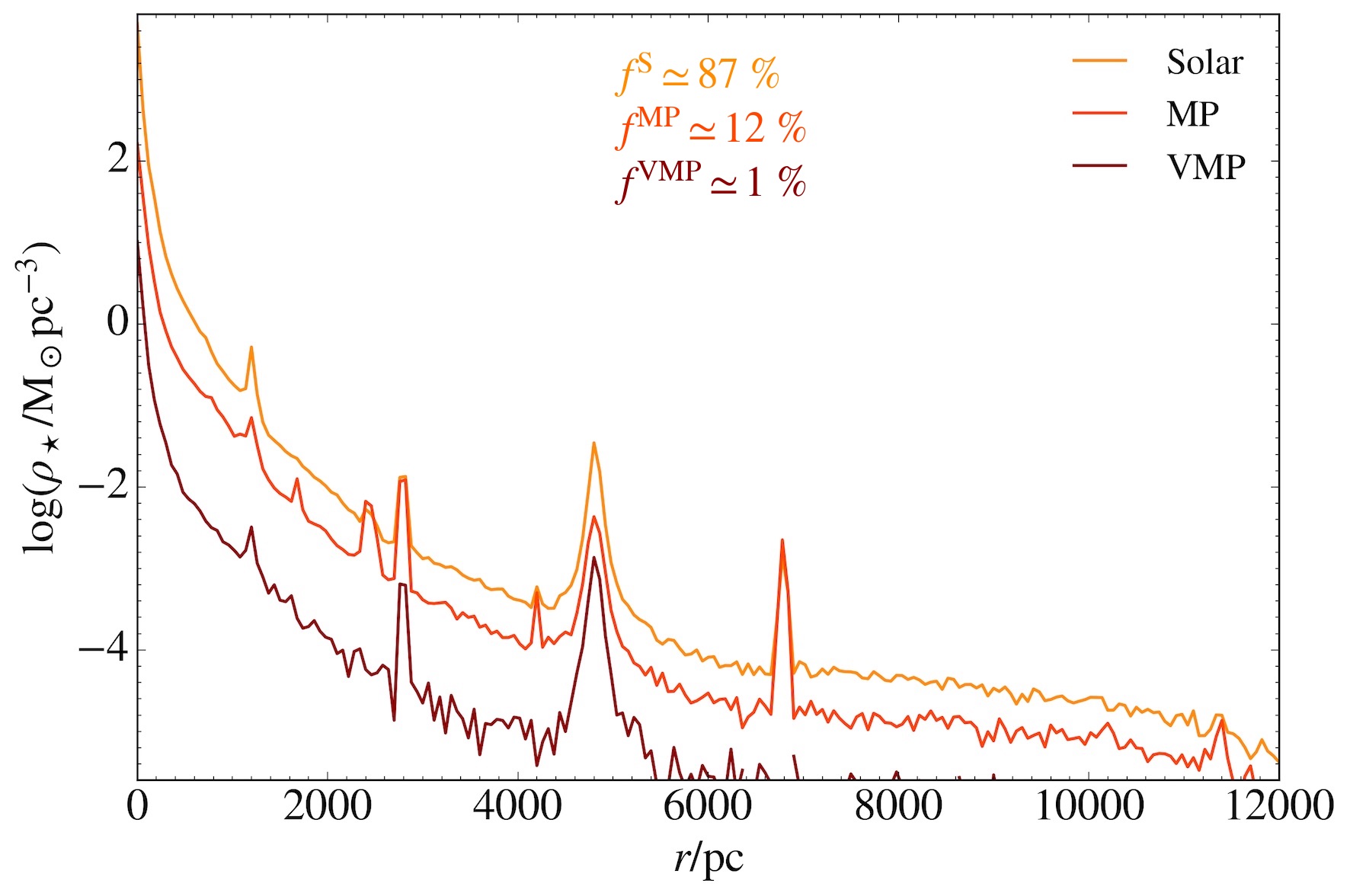}
\caption{Radial profiles of the different stellar populations of \althaea~as in Fig. \ref{fig:radial_z_bins}. \label{fig:rad_althaea_zbins} }
\end{figure}

The properties of the central LBG \althaea~fit nicely in the global trends in Fig. \ref{fig:prop_vs_mass}, as confirmed by the stellar age-metallicity relation, SFH, and MDF, all shown in Fig. \ref{fig:althaea_sfh}. The effect of merger events (vertical stripes) fuelling SF over long times is visually remarkable. \althaea~has formed stars for $\simeq$700 Myr and, despite forming its first stars in a virtually unpolluted environment, its stellar population by $z\simeq6$ is dominated by solar metallicity stars. 
The radial profiles of the different stellar populations (Fig. \ref{fig:rad_althaea_zbins}) confirm that the more massive the system considered is, the more negligible the fraction of VMP stars present, which indeed represent only the 1$\%$ of stars in \althaea.

\section{Summary and discussion}\label{sec_conclusions}

We investigated the properties of dwarf satellites of a typical high-$z$ LBG. These small systems have not been directly observed yet, but they will be one of the most exciting targets at reach of JWST. To this aim we have analysed cosmological zoom-in simulations presented in \citep{Pallottini17}. These simulations have sufficiently high mass ($\Delta m \simeq 10^4 \rm \,M_\odot$), and spatial resolution ($\Delta x \simeq 30 \,\rm pc$) to examine the inner structure of the smallest systems surrounding the central LBG called \althaea.

We identified six dwarf satellites within $\rm \lesssim 12$ kpc from \althaea's centre, whose
stellar [gas] masses are in the range $\log (M_\star/\msun) \simeq 6-9$  [$\log (M_{gas}/\msun) \simeq4.3-7.75$]. We analysed the properties of their stellar populations in terms of spatial distributions, formation history and kinematics. The main results are:

\begin{enumerate}

\item All satellites experience periods of high star formation rates, $\rm SFR \gtrsim 5 \,M_\odot yr^{-1}$, hence they are expected to produce emission lines such as the $\rm H_\alpha$ with high luminosities (e.g. see $\rm L_\alpha-SFR$ relation from \citet{Kennicutt12}). With the high sensitive NIRSpec instrument \citep{Giardino16}, JWST will be able to detect in 10 ks emission lines of sources at $z\simeq6$ with $\rm SFR\sim2\,\msunyr$, implying that our satellites will indeed be observable.\\

\item Mechanical feedback from SNe, heating/evacuating the gas, is the key physical process able to permanently (temporarily) quench the SF in the smallest (largest) satellites.\\

\item The evolution of the satellites strongly depends on their total mass. Low-mass systems ($\rm M_\star \lesssim 5\times10^8M_\odot$) have short ($\simeq 50 \rm Myr$) and simple SFHs, which become increasingly complex and long ($> 500 \rm Myr$) in more massive systems experiencing frequent merger events.\\

\item Dynamical interactions of the satellites with \althaea~likely played a role in regulating and timing the SF activity of the largest satellites only. \\

\item For all satellites SFHs, chemical composition and morphology show no dependence on the distance from the central LBG. \\

\item The chemical properties of the stellar populations of dwarf satellites also depend on their mass. High-mass systems ($M_\star > 5\times10^8\msun$) form their first stars at $z>9$ from nearly pristine gas. Lower-mass satellites form later ($z<7$), and in a circumgalactic medium already enriched by SN-driven outflows from \althaea~and its progenitors.\\

\item In all satellites, the most metal-poor stars are concentrated towards their center ($r\lesssim200$ pc) while metal-rich stars extend up to larger radii. Both components show a remarkably spherical shape. Stellar streams/arms caused by tidal effects are only observed in two closest satellites (S2 and S5).\\ 

\item The least massive satellite S1 has very low gas content and completely lacks of DM. We speculate that it might be a possible \highz proto-globular cluster candidate \citep[e.g.][]{Vanzella17}.
\end{enumerate}

From an observational point of view we saw that, given the bursty-like SFHs of the satellites, they should be bright enough for JWST to unveil their properties at \highz. For this reason we plan to make accurate prediction for their images and spectra in an upcoming work. Based on currently available observations of \highz galaxies we could expect some of our low- and intermediate-mass satellites, characterised by stellar populations that form in short efficient bursts of SF, to be detected as Lyman Alpha Emitters as suggested by \citet{haro2020differences}.

How do our findings compare with the data available for Local Group satellites?
First, in agreement with our conclusions, there are observational evidences supporting the importance of internal and external feedback processes in regulating the SFHs of satellites around the Milky Way \citep[e.g.][]{Gallart15}.
Second, we find characteristic behaviours of stars with different metallicities in galaxies of all masses, which are encountered at all epochs independently from the redshift: (i) the fraction of MP stars decreases with increasing mass of the galaxy; (ii) metal-poor (-rich) stars tend to dwell in the inner regions (to spread out), as also reported by semi-analytical models and numerical simulations for the formation of the the Milky Way and its dwarf satellites \citep{Salvadori10,Starkenburg17}.

We have pointed out that high-$z$ dwarfs may form stars at a sustained rate ($> 5\,\msunyr$). There is a marked difference with  the local dwarf spheroidal galaxy population orbiting around the Milky Way. These objects are characterised by low SFR ($< 0.01\,\msunyr$) and metallicities ($< 0.01\, \zsun$). Hence, LBG satellites may be more similar to Blue Compact Dwarfs found at the edge of the Local Group: these are gas-rich systems characterised by ongoing SF at very high-rates, ${\rm SFR} \simeq 1-10 \,\msunyr$ \citep[e.g.][]{Tolstoy09}, or possibly (i.e. S1) tidal dwarf galaxies \citep[][]{bournaud:2006,fensch2019}.

A limitation of our study is represented by the finite mass resolution of stellar (DM) particles of $\simeq10^4\msun$ ($\simeq10^5\msun$). As a consequence we cannot resolve minihaloes hosting the first stars, which in principle could influence the results for the population of low-mass satellites. However, since we are focusing our study in a biased high-density region of the cosmic web, we expect to be here in the presence of a high UV background \citep[e.g][]{Pallottini19}: an UV flux of $G =  G_0$ roughly corresponds to a background in the Lyman-Werner band of intensity $J_{LW}\simeq 40\, {\rm J}_{21}$ (with $J_{21} = 10^{-21} {\rm erg}\, {\rm s}^{-1} {\rm cm^{-2}}\,{\rm Hz^{-1}}\,{\rm sr^{-1}}$). Considering that a critical flux of $J_{LW} \sim 10^{-2} {\rm J}_{21}$ should suppress minihalo formation \citep{johnson:2013}, the central LGB is expected to suppress star formation in such small objects already at $z\sim10$ \citep[e.g.][ for the MW environment]{Graziani15}.
Therefore accounting for the presence of minihaloes would not change significantly the properties of the simulated low-mass satellites, which formed at $z<7$.
Consistently with the absence of minihaloes hosting PopIII stars \citep[e.g.][]{Wise12,OShea15}, we had to introduce a metallicity floor of $Z_{\rm floor}=10^{-3} Z_\odot$ to mimic the expected gas pre-enrichment \citep{Dave11,Wise12,Pallottini14}.
This metallicity floor however only marginally affects our results, since stars with $Z_\star < 10^{-2} Z_\odot$ represent only a tiny ($<1$\%) contribution to the stellar mass budget of the satellites. We also note that stars with $Z_\star < 10^{-3} Z_\odot$ are an extremely rare stellar population even in smallest local dwarf galaxies with $\rm M_\star < 10^7 M_\odot$ \citep[e.g.][]{Salvadori15}.

The effect of feedback by SN and stellar winds can have a dramatic impact on the satellites evolution; they might even lead to a complete SF quenching in the smallest ones. Indeed, the simulations also predict the existence of additional eight {\it starless, gas-rich, and metal-poor} satellites. In these interesting systems, located at  $D_{\rm \althaea} =8-12$ kpc, the SF was likely inhibited by UV photo-dissociation of \HH~, as predicted by \citep{SalvadoriFerrara12}.
Because of their high and dense gas content, these small satellites might be observed as Damped Ly-$\alpha$ systems \citep{dodorico:2018}, and retain in their metal-poor gas the chemical signature of very massive first stars \citep{Salvadori19}. 
It is however worth recalling that the treatment of the UV radiation field, and hence of radiative feedback, is rather coarse in the simulation. In the future we plan to improve on this aspect by using newly available simulations, part of the \code{SERRA} suite, implementing on-the-fly, multi-frequency radiative transfer. Hydrogen ionizing radiation will be also taken into account, which could lead to additional photo-evaporation effects \citep[e.g.][]{decataldo:2019}.

\section*{Acknowledgements}
We thank the anonymous referee for the useful and constructive comments.
We acknowledge support from the PRIN-MIUR2017, {\it The quest for the first stars}, prot. n. 2017T4ARJ5.
VG and SS acknowledge support from the ERC Starting Grant NEFERTITI H2020/808240.
AF acknowledge support from the ERC Advanced Grant INTERSTELLAR H2020/740120.
Any dissemination of results must indicate that it reflects only the author’s view and that the Commission is not responsible for any use that may be made of the information it contains. 
Support from the Carl Friedrich von Siemens-Forschungspreis der Alexander von Humboldt-Stiftung Research Award is kindly acknowledged (AF).
We thank \'Asa Skuladottir for careful reading and comments of the draft and Suzanne Madden for interesting discussions.
We acknowledge use of the Python programming language \citep{VanRossum1991}, Astropy \citep{astropy}, Matplotlib \citep{Hunter2007}, NumPy \citep{VanDerWalt2011}, \code{pynbody} \citep{pynbody}, and SciPy \citep{scipyref}. \\
Data Availability Statement: the data underlying this article were accessed from the computational resources available to the Cosmology Group at Scuola Normale Superiore, Pisa (IT). The derived data generated in this research will be shared on reasonable request to the corresponding author.




\bibliographystyle{mnras}
\bibliography{refer,codes} 


\appendix
\section{Stellar streams}\label{app:streams}

In this Appendix we discuss the origin of the stellar streams that are present in the external regions of the two satellites S2 and S5 (see Sec. \ref{subsec_densmaps}). The presence of these streams may be expected since they are the two closest satellites to \althaea~($D_{\rm \althaea} < 2.5$ kpc), and therefore are most likely affected by tidal interactions. We here study the stellar kinematics in these two satellites to interpret the origin of these structures.

\begin{figure}
\centering
\includegraphics[width=0.5\textwidth]{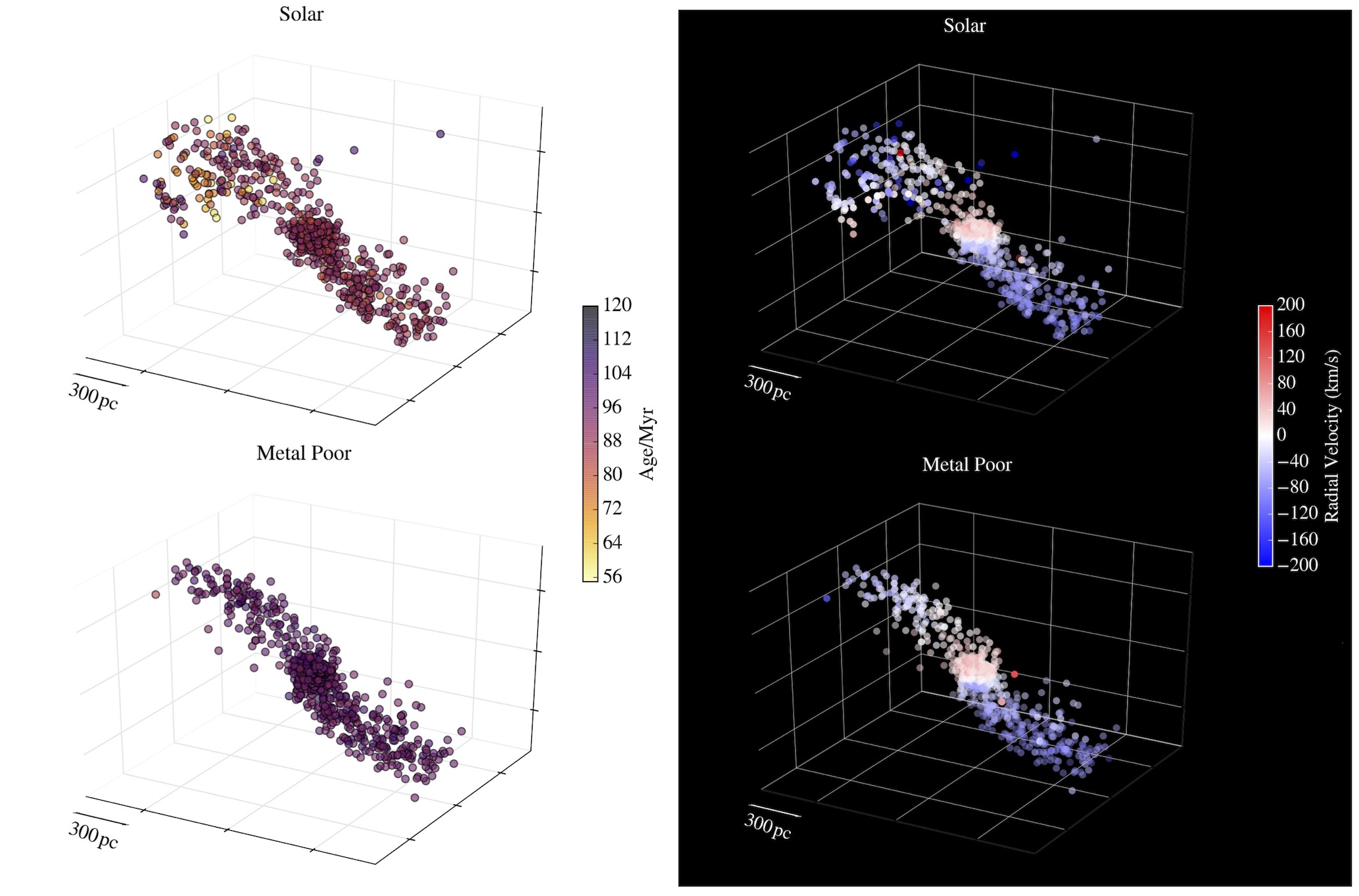}
\caption{3D spatial maps of stars in satellite S2. 
The star particles have been divided in metallicity intervals as in Fig. \ref{fig:radial_z_bins}. The colorbar in the left column shows the stellar ages, while the one in the right column shows the radial velocities of stars with respect to the centre of mass of the satellite.
\label{fig:S2_3D}
}
\end{figure}

\begin{figure}
\centering
\includegraphics[width=0.5\textwidth]{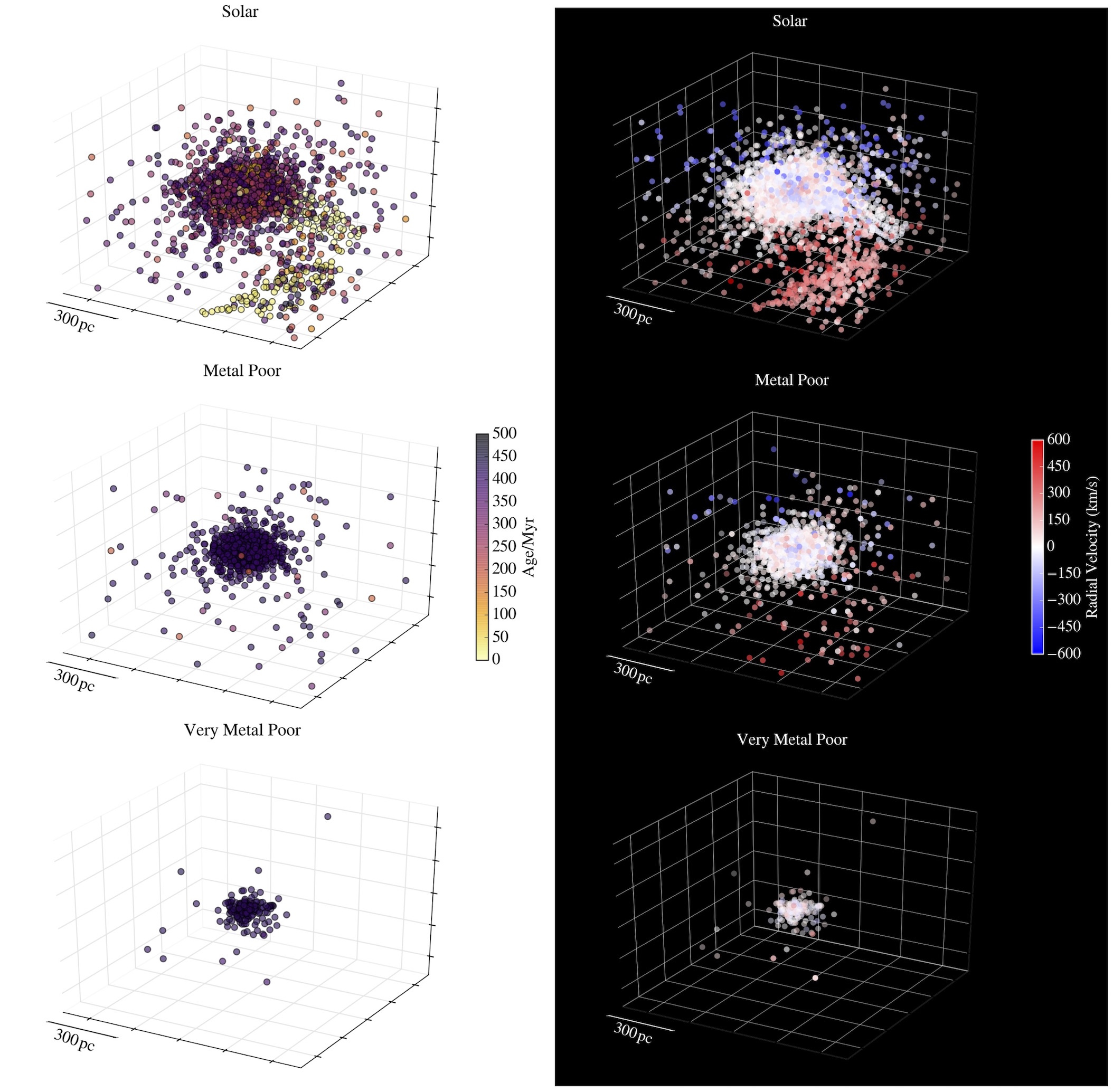}
\caption{3D maps of satellite S5, as in Fig. \ref{fig:S2_3D}.
\label{fig:S5_3D}
} 
\end{figure}

In Fig. \ref{fig:S2_3D} and \ref{fig:S5_3D} we display the spatial distribution of the stellar populations of S2 and S5 respectively (using the same metallicity bins as in Sec. \ref{stellarmetallicity}). The colours in the left column indicate the stellar age and those in the right indicate the stellar radial velocity with respect to the centre of mass of the satellite.

We find that in S2 the streams are moving toward the centre of the satellite at high velocities,  $\rm v \simeq$ 100 km/s, which is greater than the mean velocity of the stars into the bulge, $\rm \langle v\rangle \simeq$ 50 km/s. 
We observe that the stellar streams are composed by both Solar and MP stars, reflecting the same chemical composition of the stellar populations found in the central spherical bulge. This implies that all these stars, both in the streams and in the bulge, are most likely born in the same dwarf galaxy (over a relatively brief period of time $\simeq50$ Myr), and that the formation of the streams was caused by a subsequent tidal interaction with \althaea.
The most plausible scenario is the one in which the satellite have interacted with the disk of the nearby massive LBG while rotating. Indeed, the rotation likely caused two arms to detach, and at redshift $z\simeq6$ the stars inside them are falling back onto the bulge.

In S5 the stellar particles forming the single stream are found in the Solar metallicity population only. These stars are all very young and they are moving away from the satellite at velocity higher than the escape velocity, $\rm v_{esc} \simeq 220 \rm km/s$. They are thus falling directly onto \althaea, subjected to its strong gravitational pull.
Recalling Fig. \ref{fig:mapdens}, we notice that also the gas follows a similar spatial distribution as stars, and an analogous infalling stream can in fact be found. This suggests that, since the stellar stream is composed by very young stars, these stars may have formed within the gas rich stream itself while infalling onto \althaea.

We conclude that the streams found in the two satellites, even though caused in both cases by the presence of the nearby \althaea, have a different nature: in S2 we are witnessing a perturbation of the spatial distribution of stars pertaining to a already formed dwarf galaxy; in S5 with a population of newly formed stars moving from the dwarf satellite to the central LBG \althaea.

\bsp	
\label{lastpage}
\end{document}